\documentclass[12pt]{article}%
\usepackage{amsfonts}
\usepackage{amsmath}
\usepackage{amssymb}
\usepackage{graphicx}
\usepackage[superscript]{cite}%
\providecommand{\U}[1]{\protect \rule{.1in}{.1in}}
\providecommand{\U}[1]{\protect \rule{.1in}{.1in}}
\providecommand{\U}[1]{\protect \rule{.1in}{.1in}}

\def \doublespace {\openup 2.0\jot}
\doublespace
\begin{document}

\title{On Semiparametric Efficiency of an Emerging Class of Regression Models for
Between-subject Attributes}

\author{Jinyuan Liu$^{1}$ (jil1168@health.ucsd.edu),\\ \and Tuo Lin$^{1}$ (tulin@health.ucsd.edu),\\ \and Tian Chen$^{2}$ (tian.chen1@takeda.com),\\ \and Xinlian Zhang$^{*1}$ (xizhang@health.ucsd.edu),\\ \and Xin M. Tu$^{1}$ (x2tu@health.ucsd.edu)\\ \and $^{1}$Department of Family Medicine and Public Health\\UC San Diego, San Diego, CA \\$^{2}$Takeda Pharmaceuticals,
Cambridge, MA}

\maketitle

\begin{center}
\textbf{Abstract}
\end{center}

The semiparametric regression models have attracted increasing attention owing
to their robustness compared to their parametric counterparts. This paper
discusses the efficiency bound for functional response models (FRM), an
emerging class of semiparametric regression that serves as a timely solution
for research questions involving pairwise observations. This new paradigm is
especially appealing to reduce astronomical data dimensions for those arising
from wearable devices and high-throughput technology, such as microbiome
Beta-diversity, viral genetic linkage, single-cell RNA sequencing, etc.
Despite the growing applications, the efficiency of their estimators has not
been investigated carefully due to the extreme difficulty to address the
inherent correlations among pairs. Leveraging the Hilbert-space-based
semiparametric efficiency theory for classical within-subject attributes, this
manuscript extends such asymptotic efficiency into the broader regression
involving between-subject attributes and pinpoints the most efficient
estimator, which leads to a sensitive signal-detection in practice. With
pairwise outcomes burgeoning immensely as effective dimension-reduction
summaries, the established theory will not only fill the critical gap in
identifying the most efficient semiparametric estimator but also propel
wide-ranging implementations of this new paradigm for between-subject attributes.


\newpage

\section{Introduction\label{Ef.sec1}}

As a mainstream, the classical generalized linear model (GLM)
\cite{Agresti0000} encompasses nonnormal and noncontinuous responses (or
dependent variables) to present a unified paradigm for different response
types. The maximum likelihood estimators (MLE) for GLM enjoy consistency and
asymptotic normality (CAN) if both the random and systematic components, as
well as the link function, are correctly specified. Estimators with
(asymptotic) variances achieving the Cram\.{e}r-Rao bound are referred to as
(asymptotically) efficient; MLEs are exact examples of efficient estimators
for the parametric GLM.

By relaxing the nuisance parameter in the distributional assumption to be
infinite-dimensional, the semiparametric GLM, also termed the restricted
moment models (RMM) \cite{Tsaitis2006}$,$ enables robust statistical inference
for a broader class of data distributions. Under suitable regularity
conditions, estimators from the generalized estimating equations (GEE) are
optimal. They enjoy consistency and asymptotically normality \cite{Zeger} and
also achieve the semiparametric efficiency bound \cite{Tsaitis2006}$.$

Despite its immense applicability, GLM predominantly focuses on the
relationships within the same subject, termed \textquotedblleft within-subject
attributes\textquotedblright. But in the growing applications, of major
interest are outcomes defined by a pair of subjects, or the \textquotedblleft
between-subject attributes$.$\textquotedblright \  \cite{Liu2021} The
probability index $Pr(Y_{i_{1}}<Y_{i_{2}}),$ $(i_{1},i_{2})\in C_{2}^{n}$ in
the Mann-Whitney-Wilcoxon (MWW) rank-sum test is a classical example
\cite{chen2016}$.$ Fueled by innovative technologies such as high-throughput
sequencing and wearable devices, the pairwise dissimilarity/distance metrics
that summarize high-dimensional sequences also entail a between-subject nature
\cite{nguyen2021}$.$

Modeling between-subject attributes are challenging due to the complex
correlation structures among pairwise observations. To address this, a
semiparametric framework of functional response models (FRM) has been
proposed, which naturally enlarges GLM to involve between-subject attributes
\cite{Liu2}$.$ This paradigm accommodates various types of between-subject
distances and complements the prevailing regularization-based approach for
high-dimensional data.

The FRM framework admits a wide range of applications. For instance, Liu et
al. (2021) \cite{Liu2021} adopted the FRM to extend the predominant
ANOVA-based approach to model microbiome Beta-diversity in a regression; Lin
et al. (2021) \cite{lin2021} implemented the MWW rank-sum test and FRM in
survey data to address the restrictive test of equal distributions; Wu et al.
(2014) \cite{wu2014} incorporated the inverse probability weighting (IPW) into
a rank-based statistic and applied FRM to deal with confounding effects in
causal inference. Their estimators from the U-statistics-based generalized
estimating equations (UGEE) \cite{Kowalski2007} all enjoy nice asymptotic
properties just like their GEE counterparts for the semiparametric GLM.

Nevertheless, the efficiency of UGEE estimators for the semiparametric FRM has
not been investigated thoroughly. Akin to the case for within-subject
attributes, we aim to find estimator(s) with the smallest asymptotic variance,
or the semiparametric efficient estimator(s) for between-subject attributes.
To this end, one first needs to extend essential concepts (such as influence
functions and asymptotic linearity) from the classical within-subject settings
and then develop a coherent theory in the FRM regression for between-subject
attributes. In this manuscript, we leverage the Hilbert-space-based
semiparametric efficiency theory to demonstrate that the UGEE estimators also
achieve the efficiency bound, just like GEE estimators for the semiparametric
GLM. Hence, harmonizing the semiparametric efficiency and robustness, the
modeling framework for between-subject attributes can facilitate knowledge
discovery for scientific questions that call for such models and inform
appropriate decision making.

The rest of the manuscript is organized as follows. We first introduce the FRM
framework (with examples) in Section 2 and fundamentals for semiparametric
efficiency in Section 3. We then generalize the Hilbert space tailored for
between-subject attributes in Section 4. In Sections 5, 6, and 7, we show the
semiparametric efficiency of UGEE estimators through a \textquotedblleft
conjugate" class of models, leveraging geometric perspectives built upon
Hilbert spaces. Examples of efficient UGEE estimators are then demonstrated in
Section 8. We also discuss adaptive estimators and present simulation studies
in Section 9. In section 10, we give our concluding remarks.

\section{Between-subject Functional Response Models \label{Ef.sec2}}

\subsection{Between-subject Attributes\label{Ef.sec2.0}}

Fueled by technological advances such as next-generation sequencing and
wearable devices, between-subject attributes are gaining popularity by
reducing dimensions effectively. They have evolved into the center stage of
biomedical and other burgeoning research areas, such as microbiome,
single-cell RNA sequencing \cite{moon2017phate}$,$ etc. To illustrate, the
human microbiome is now interrogated using high-throughput sequencing (e.g.,
16s sequencing of gut microbiota) for insights in disease mechanisms. This
procedure generates taxonomic sequence counts (for each subject) that are
sparse and astronomically high-dimensional (e.g., in our data application, the
dimension $m=12,131$). Due to their additional sparsity and non-normality, the
microbiome \textquotedblleft diversity" has been introduced to summarize the
raw sequence. This biologically-relevant concept constitutes a critical
indicator of human health \cite{lozu2012}$.$

For example, the Beta-diversity defined by the pairwise distance of taxonomic
sequence counts naturally encompasses a between-subject nature. Consider a
human microbiome dataset composed of $n$ subjects. Let $\mathbf{Y}_{i}$ $\in%
%TCIMACRO{\U{211d} }%
%BeginExpansion
\mathbb{R}
%EndExpansion
^{m}$ denote a column vector of relative abundance (proportions) of taxonomic
units for the $i$-th subject, the Aitchison Beta-diversity
\cite{Aichinson1982} between any pair $\left(  i_{1},i_{2}\right)  \in
C_{2}^{n}$ is\  \  \
\begin{equation}
d_{A}(\mathbf{Y}_{i_{1}},\mathbf{Y}_{i_{2}})=\left[
%TCIMACRO{\dsum \limits_{k=1}^{m}}%
%BeginExpansion
{\displaystyle \sum \limits_{k=1}^{m}}
%EndExpansion
\left(  \log \frac{Y_{i_{1}k}}{g(\mathbf{Y}_{i_{1}})}-\log \frac{Y_{i_{2}k}%
}{g(\mathbf{Y}_{i_{2}})}\right)  ^{2}\right]  ^{1/2},\text{ }g(\mathbf{Y}%
_{i})=\left(
%TCIMACRO{\dprod \limits_{k=1}^{m}}%
%BeginExpansion
{\displaystyle \prod \limits_{k=1}^{m}}
%EndExpansion
Y_{ik}\right)  ^{^{1/m}}, \label{eqn.n1}%
\end{equation}
where $C_{q}^{n}$ denotes the set of $q$-combinations $\left(  i_{1}%
,\ldots,i_{q}\right)  $ from the integer set $\left \{  1,\ldots,n\right \}  $,
$g(Y_{i})$ is the geometric mean of $Y_{i}$.

By integrating information from the raw high-dimensional sequences,
Beta-diversity measures the dissimilarity/distance between two subjects across
all (or a proportion of) the sequenced genomes thus merits its scientific interest.

The versatility of between-subject attributes motivates researchers to migrate
such pairwise distances to the blooming real-time longitudinal sequences
collected from wearables. For example, since the mean of squared Euclidean
distance pertains to the variance, pairwise distances of those
high-dimensional sequences can naturally capture the between-subject
variability (beyond the population mean). By unraveling the intricate
connections between physical activity and clinical traits, between-subject
distances could facilitate personalized disease interventions.

Another example of between-subject attribute is the connection between two
subjects in a social network. Since the connection is defined for more than
one subject, the subject-level outcome is latent here.

For the examples above, main interests are shifted to the between-subject
attributes (instead of the raw high-dimensional sequences). We refer to them
as \textquotedblleft endogenous\textquotedblright \ to distinguish them from
another category called \textquotedblleft exogenous\textquotedblright%
\ between-subject attributes, where the focus is still on the original
within-subject attributes. For instance, for a scalar within-subject $Y_{i}$,
the squared difference index $(Y_{i_{1}}-Y_{i_{2}})^{2}$ can extend the ANOVA
to compare variances (rather than means) among groups \cite{zhang2020}$,$ and
a probability index $Pr(Y_{i_{1}}<Y_{i_{2}})$ can compare groups in the MWW
rank-sum test \cite{wu2014} to address outliers. In these examples, we are
interested in characteristics of the original within-subject attributes, such
as variances or differences between two distributions.

The distinction between these two types of between-subject attributes is not
as rigorous, but differentiating them will enable us to handle the raw data
more systematically during statistical modeling.

\subsection{Semiparametric GLM and Functional Response Model\label{sec3}}

Consider a study with $n$ subjects, let $Y_{i}$ denote a response, $X_{i}$ an
explanatory variable for the $i$-th subject. As a motivating example, the
semiparametric GLM (SPGLM) characterizing the relationship between $Y_{i}$ and
$X$$_{i}$ is:
\begin{equation}
E\left(  Y_{i}\mid X_{i}\right)  =h(X_{i};\boldsymbol{\beta}),\text{ }1\leq
i\leq n, \label{eqn2.00}%
\end{equation}
where $h\left(  \cdot \right)  $ is the inverse of some link functions
\cite{Tang2011}$,$ additional explanatory variables can be added to the linear
predictor. Compared with the classical parametric GLM, (\ref{eqn2.00}) is more
flexible by removing the distributional assumption on $Y_{i}$ thus yields
valid inference even when the data deviate from such an assumption. However,
limitations of this prevalent framework include: 1) it does not apply to the
between-subject, or pairwise, attributes that are of interest in a mounting
number of applications; 2) it fails to directly model a multivariate response
$\mathbf{Y}_{i}\in%
%TCIMACRO{\U{211d} }%
%BeginExpansion
\mathbb{R}
%EndExpansion
^{m}$ ($m\geq1$), especially when the dimension of $\mathbf{Y}_{i}$ is high.

Hence, we adopt an enlarged paradigm. Consider observing the raw data
$(\mathbf{Y}_{i}^{\top},\mathbf{X}_{i}^{\top})$, where $\mathbf{Y}%
_{i}(\mathbf{X}_{i})\in%
%TCIMACRO{\U{211d} }%
%BeginExpansion
\mathbb{R}
%EndExpansion
^{m}$ ($m\geq1$) is a column vector of multivariate response (explanatory
variable) for the $i$-th subject. By concatenating $\mathbf{Y}_{i}$ into a
(scalar) functional response of multiple ($s$) subjects $f\left(
{\normalsize \mathbf{Y}}_{i_{1}},\ldots,{\normalsize \mathbf{Y}}_{i_{s}%
}\right)  $, $\left(  i_{1},\ldots,i_{s}\right)  \in C_{s}^{n},$ the
semiparametric framework of functional response models (FRM) resolves the
aforementioned challenges:%

\begin{equation}
E\left[  f\left(  \mathbf{Y}_{i_{1}},\ldots,\mathbf{Y}_{i_{s}}\right)
\mid \mathbf{X}_{i_{1}},\ldots,\mathbf{X}_{i_{s}}\right]  =h\left(
\mathbf{X}_{i_{1}},\ldots,\mathbf{X}_{i_{s}};\boldsymbol{\beta}\right)
,\text{ }\left(  i_{1},\ldots,i_{s}\right)  \in C_{s}^{n},\text{ }s\geq1,
\label{eqn.2.10.new}%
\end{equation}
where $f(\cdot)$ is some scalar-valued function, $h\left(  \cdot \right)  $ is
some smooth function (e.g., with continuous derivatives up to the second
order), $\beta$\ is a vector of parameters, $s$ is a positive integer. Akin to
(\ref{eqn2.00}), (\ref{eqn.2.10.new}) is also semiparametric without any
distributional assumption on the response $f\left(  {\normalsize \mathbf{Y}%
}_{i_{1}},\ldots,{\normalsize \mathbf{Y}}_{i_{s}}\right)  $. In practice, this
introduces greater flexibility, which addresses the difficulty to specify such
an assumption for the multi-subject based response function $f\left(
{\normalsize \mathbf{Y}}_{i_{1}},\ldots,{\normalsize \mathbf{Y}}_{i_{s}%
}\right)  $ that resembles the real study data. As a special case when $m=s=1$
and $f\left(  \mathbf{Y}_{i}\right)  =Y_{i}$, (\ref{eqn.2.10.new})\ reduces to
(\ref{eqn2.00}). The FRM is also readily extended to model a vector-valued
response function (see Example 3 in Section \ref{Ef.sec2.1}).

We now implement (\ref{eqn.2.10.new})\ to model the between-subject
attributes, which will be the focus of this paper. For notational consistency,
we use $i$ to index a subject and $\mathbf{i}=\left(  i_{1},i_{2}\right)  \in
C_{2}^{n}$ to index a pair in what follows. Let $s=2$, we construct a
between-subject attribute $f_{\mathbf{i}}=f\left(  \mathbf{Y}_{i_{1}%
},\mathbf{Y}_{i_{2}}\right)  $ with some mapping such as (\ref{eqn.n1}), then
the semiparametric FRM\ below models $f_{\mathbf{i}}$ as a function of
$\mathbf{X}_{\mathbf{i}}=\left(  \mathbf{X}_{i_{1}}^{\top},\mathbf{X}_{i_{2}%
}^{\top}\right)  ^{\top}:$ \
\begin{equation}
E\left(  f_{\mathbf{i}}\mid \mathbf{X}_{\mathbf{i}}\right)  =h\left(
\mathbf{X}_{\mathbf{i}};\boldsymbol{\beta}\right)  ,\text{ }\mathbf{i}=\left(
i_{1},i_{2}\right)  \in C_{2}^{n}. \label{eqn 2.15}%
\end{equation}

(\ref{eqn 2.15}) extends the classical GLM from within- to between-subject
attributes that are generally tricky to model. It not only achieves effective
dimension-reduction but also establishes a complementing angle for data
entailing an intrinsic between-subject nature. Hence, semiparametric FRM
uniquely positions itself to facilitate data-driven knowledge discoveries,
which could otherwise be hindered by the predominant paradigm of merely
modeling within-subject attributes. We highlight the widespread applicability
of semiparametric FRM with additional examples below. More applications can be
found in the references of Section \ref{Ef.sec1}. \ 

\subsection{Examples of Functional Response Models\label{Ef.sec2.1}}

\textbf{Example 1: The Beta-diversity for High-throughput Data in Microbiome}

For many studies such as the microbiome, major interest is to compare
characteristics of the between-subject attributes among subgroups, where FRM
is desirable. We start with a categorical variable $X_{i}$ with $K$ levels
(such as the disease status). To transform $X_{i}$ to a between-subject
attribute for the $i$-th pair, we define a set of pairwise indicators (or
dummy variables) for $X_{\mathbf{i}}=\left \{  X_{i_{1}},X_{i_{2}}\right \}  $
through the one-hot encoding\ function $\delta \left(  \mathbf{\cdot}\right)
:\{1,...,K\} \times \{1,...,K\} \mapsto \{0,1\}^{K+C_{2}^{K}}:$
\begin{align}
\delta_{k_{1}k_{2}}\left(  X_{\mathbf{i}}\right)   &  =\left \{
\begin{array}
[c]{ll}%
1, & \text{if }X_{\mathbf{i}}=\left \{  X_{i_{1}},X_{i_{2}}\right \}  =\left \{
k_{1},k_{2}\right \}  \text{ }\mathbf{,}\\
0, & \text{otherwise.}%
\end{array}
\right. \label{eqn3.60}\\
\boldsymbol{\delta}\left(  X_{\mathbf{i}}\right)   &  =\left(  \delta
_{11}\left(  X_{\mathbf{i}}\right)  ,\ldots,\delta_{\left(  K-1\right)
K}\left(  X_{\mathbf{i}}\right)  ,\delta_{KK}\left(  X_{\mathbf{i}}\right)
\right)  ^{\top},\quad1\leq k_{1}\leq k_{2}\leq K.\nonumber
\end{align}
where the vector $\boldsymbol{\delta}(X_{\mathbf{i}})\in$ $%
%TCIMACRO{\U{211d} }%
%BeginExpansion
\mathbb{R}
%EndExpansion
^{K+C_{2}^{K}}$ denotes all combinations. Thus, $\delta_{k_{1}k_{2}}\left(
X_{\mathbf{i}}\right)  $ indicates the pair with the same $k$-th concordant
($k_{1}=k_{2}=k$) or discordant\ ($k_{1}<k_{2}$) levels for $X_{\mathbf{i}}$.

For example, if $X_{i}$ is a binary indicator of disease, we form
\[
\boldsymbol{\delta}\left(  X_{\mathbf{i}}\right)  =\left(  \delta_{DD}\left(
X_{\mathbf{i}}\right)  ,\delta_{HH}\left(  X_{\mathbf{i}}\right)  ,\delta
_{HD}\left(  X_{\mathbf{i}}\right)  \right)  ^{\top},
\]
where $\delta_{DD}\left(  X_{\mathbf{i}}\right)  $ and $\delta_{HH}\left(
X_{\mathbf{i}}\right)  $ index diseased-diseased and healthy-healthy pairs,
and $\delta_{HD}\left(  X_{\mathbf{i}}\right)  $ represents the mixed
healthy-diseased pairs.

Let $f_{\mathbf{i}}=f\left(  \mathbf{Y}_{i_{1}},\mathbf{Y}_{i_{2}}\right)  $
denote the Beta-diversity for the $i$-th pair such as the Aitchison
distance\ in\ (\ref{eqn.n1}), we can model its mean among subgroups adopting
the FRM:
\begin{equation}
E\left(  f_{\mathbf{i}}\mid \mathbf{X}_{\mathbf{i}}\right)  =\exp \left[
\boldsymbol{\beta}^{\top}\boldsymbol{\delta}(X\mathbf{_{\mathbf{i}}})\right]
,\text{ }\boldsymbol{\beta}=\left(  \tau_{11},\ldots,\tau_{KK}\right)  ^{\top
},\text{ }\boldsymbol{\delta}\left(  X_{\mathbf{i}}\right)  =\left(
\delta_{11}\left(  X_{\mathbf{i}}\right)  ,\ldots,\delta_{KK}\left(
X_{\mathbf{i}}\right)  \right)  ^{\top}, \label{eqn.110}%
\end{equation}
where $\exp \left(  \cdot \right)  $ ensures that the response is non-negative.

The coefficients of the dummy variables now reveal the heterogeneity in
$f_{\mathbf{i}}$ among different subgroups defined by $\boldsymbol{\delta
}\left(  X_{\mathbf{i}}\right)  $. Such a pairwise one-hot encode also
facilitates disentangling different types of heterogeneity (e.g.,
\textquotedblleft location\textquotedblright \ or \textquotedblleft
scale\textquotedblright \ difference) \cite{Liu2021}$,$ which is laborious or
not even feasible using existing approaches such as PERMANOVA. We can also
include either between- or within-subject attributes as covariates in
(\ref{eqn.110}). For between-subject covariates, it is straightforward. For
within-subject attributes, we can readily create their between-subject
counterparts \cite{Liu2021} as shown above. \ 

\textbf{Example 2: Mann-Whitney-Wilcoxon Rank-sum Test and Rank Regression }

Let $Y_{i}$ ($1\leq i\leq n$) denote a univariate continuous within-subject
response. In the presence of outliers, the Mann-Whitney-Wilcoxon rank-sum test
offers a robust alternative to the two-sample t-test to compare the centers of
two distributions \cite{lin2021}$.$ Let $\mathbf{X}_{i}$ denote a vector of
explanatory variables for the $i$-th subject, FRM readily extends the rank-sum
test to a regression:
\begin{equation}
E\left(  f_{\mathbf{i}}\mid \mathbf{X}_{\mathbf{i}}\right)  =h\left(
\mathbf{X}_{\mathbf{i}};\boldsymbol{\beta}\right)  =\Phi \left[
-\boldsymbol{\beta}^{\top}\mathbf{(X}_{i_{1}}-\mathbf{X}_{i_{2}}%
\mathbf{)}\right]  ,\text{ }f_{\mathbf{i}}=f\left(  Y_{i_{1}},Y_{i_{2}%
}\right)  =I\left(  Y_{i_{1}}\leq Y_{i_{2}}\right)  , \label{eqn.201.n}%
\end{equation}
where $\Phi \left(  \cdot \right)  $ denotes the cumulative distribution
function (CDF) of the standard normal distribution. The parameter
$\boldsymbol{\beta}$ in (\ref{eqn.201.n}) preserves its interpretation in the
conventional linear model for within-subject attributes by regressing $Y_{i}$
on $\mathbf{X}_{i}$ but considerably addresses outliers in $Y_{i}$. Unlike
Example 1, in this exogenous example, research interest still centers on the
relationship between $Y_{i}$ and $\mathbf{X}_{i}$. It was further extended to
longitudinal settings with missing values \cite{chen2016}$.$ More examples can
be found in literatures for probability index models \cite{JRSSB}$.$

\textbf{Example 3: Intraclass Correlations for Rater Agreement }

Consider a study of $n$ subjects in which each subject is rated by $K$ judges.
Let $Y_{ik}$ denote the rating for the $i$-th subject by the $k$-th judge
($1\leq k\leq K$), it is commonly characterized by\ a two-way mixed-effects
model:
\begin{align}
Y_{ik}  &  =\mu+\beta_{i}+\gamma_{k}+\left(  \beta \gamma \right)
_{ik}+\varepsilon_{ik},\text{ }\varepsilon_{ik}\sim N\left(  0,\sigma
_{\varepsilon}^{2}\right)  ,\label{eqn.2.50}\\
\beta_{i}  &  \sim N\left(  0,\sigma_{\beta}^{2}\right)  ,\text{ }\sum
_{k=1}^{K}\gamma_{k}=0,\text{ }\left(  \beta \gamma \right)  _{ik}\sim N\left(
0,\sigma_{\beta \gamma}^{2}\right)  ,\text{ }\sum_{k=1}^{K}\left(  \beta
\gamma \right)  _{ik}=0,\nonumber
\end{align}
where $N\left(  0,\sigma^{2}\right)  $ denotes a normal distribution with mean
$0$ and variance $\sigma^{2}$. If interest is in the agreement among $K$
judges, a widely applied index is the intraclass correlation (ICC)
$\rho=\left[  \sigma_{\beta}^{2}-\sigma_{\beta \gamma}^{2}/(K-1)\right]
/\left(  \sigma_{\beta}^{2}+\sigma_{\beta \gamma}^{2}+\sigma^{2}\right)  $
\  \cite{shrout1979intraclass}$,$ which can be computed after fitting the model
in (\ref{eqn.2.50}). But the major concern is the difficulty to validate the
multiple imposed normal assumptions, especially for the random effects
$\left(  \beta \gamma \right)  _{ik}$ due to their latent nature. Hence, it
renders the likelihood-based approaches prone to invalid inference under the
rating data that is usually non-normal.

This is readily fixed with a semiparametric alternative. Let
\begin{align*}
\overline{Y}_{i\cdot}  &  =\frac{1}{K}\sum_{k=1}^{K}Y_{ik}\mathbf{,}\text{
}f_{\mathbf{i}1}=\frac{1}{2}\left(  \overline{Y}_{i_{1}\cdot}-\overline
{Y}_{i_{2}\cdot}\right)  ^{2}\mathbf{,}\text{ }g_{\mathbf{i}k}=\frac{1}%
{2}\left(  Y_{i_{1}k}-Y_{i_{2}k}\right)  ^{2}\mathbf{,}\\
f_{\mathbf{i}2}  &  =\frac{1}{K}\sum_{k=1}^{K}g_{\mathbf{i}k},\text{
}h_{\mathbf{i}1}=\frac{\left[  1+\left(  K-1\right)  \rho \right]  \tau^{2}}%
{K},\text{ }h_{\mathbf{i}2}=\tau^{2},
\end{align*}
we construct the following (multivariate) FRM:%
\begin{equation}
E\left(  \mathbf{f}_{\mathbf{i}}\right)  =\mathbf{h}_{\mathbf{i}}\left(
\boldsymbol{\theta}\right)  ,\text{ }\mathbf{f}_{\mathbf{i}}=\left(
f_{\mathbf{i}1},f_{\mathbf{i}2}\right)  ^{\top},\text{ }\mathbf{h}%
_{\mathbf{i}}=\left(  h_{\mathbf{i}1},h_{\mathbf{i}2}\right)  ^{\top},\text{
}\boldsymbol{\theta}=\left(  \tau^{2},\rho \right)  ^{\top}. \label{eqn2.801}%
\end{equation}
The $\rho$ in (\ref{eqn2.801}) is exactly the ICC \cite{lu2014} that we aim to
model. In addition to robustness, this model also allows for an immediate
extension to longitudinal settings.

\subsection{Inference for the U-statistics and UGEE \label{Ef.sec2.3}}

The FRM reinvigorates regression by extending within- to between-subject
attributes. However, popular asymptotic methods such as central limit theorem
(CLT) rely on the critical assumption of independence and as such are not
directly applicable to FRM, since the functional responses\ are correlated.
This is addressed via the theory of U-statistics \cite{hoeffding1948}.

\subsubsection{Asymptotic Properties of U-statistics}

Most statistics in modeling within-subject attributes are summations of
$i.i.d.$ elements, such as the score and estimating equations. However,
statistics formed by between-subject attributes in the FRM are correlated. To
resolve this issue, a class of U-statistics-based generalized estimating
equations (UGEE) have been developed. We briefly review the U-statistics that
are instrumental in studying multi-subject-based statistics.

\textbf{Definition.} \ Consider a sample of $i.i.d.$ random vectors
$\mathbf{Y}_{i}\in%
%TCIMACRO{\U{211d} }%
%BeginExpansion
\mathbb{R}
%EndExpansion
^{m}$ ($1\leq i\leq n$). Let $d^{d\times1}\left(  \mathbf{Y}_{1}%
,\ldots,\mathbf{Y}_{s}\right)  $ be a $d$-dimensional symmetric function with
$s$ input vectors (or arguments), i.e., $d\left(  \mathbf{Y}_{1}%
,\ldots,\mathbf{Y}_{s}\right)  =$ $d\left(  \mathbf{Y}_{i_{1}},\ldots
,\mathbf{Y}_{i_{s}}\right)  $ for any permutation $\left(  i_{1},\ldots
,i_{s}\right)  $ of $\left(  1,\ldots,s\right)  $. A $d$-variate, one-sample,
$s$-argument U-statistic is \
\begin{equation}
\mathbf{U}_{n}=\binom{n}{s}^{-1}\sum_{\left(  i_{1},\ldots,i_{s}\right)
\mathbf{\in}C_{s}^{n}}\mathbf{f}\left(  \mathbf{Y}_{i_{1}},\ldots
,\mathbf{Y}_{i_{s}}\right)  ,\text{ }s\geq1, \label{U.1}%
\end{equation}
where $C_{s}^{n}=\{ \left(  i_{1},\ldots,i_{s}\right)  ;1\leq i_{1}%
<\ldots<i_{s}\leq n\}$ denotes the set of all distinct $s$-combinations from
the integer set $\left \{  1,\ldots,n\right \}  $. It is easily checked that
$E\left(  \mathbf{U}_{n}\right)  =E\left[  \mathbf{f}\left(  \mathbf{Y}%
_{i_{1}},\ldots,\mathbf{Y}_{i_{s}}\right)  \right]  =\boldsymbol{\theta}$,
i.e., $\mathbf{U}_{n}$ is an unbiased estimator of $\boldsymbol{\theta}$.

Since $f\left(  \mathbf{Y}_{i_{1}},\ldots,\mathbf{Y}_{i_{s}}\right)  $ (also
termed the kernel function) involves multiple rather than a single subject,
dependencies between any two kernel functions arise when they share at least
one common subject (e.g., $f\left(  \mathbf{Y}_{i_{1}},\mathbf{Y}_{i_{2}%
}\right)  $ and $f\left(  \mathbf{Y}_{i_{1}},\mathbf{Y}_{i_{3}}\right)  $ are
correlated as they share $\mathbf{Y}_{i_{1}}$). This dependency is tackled
through the $H\acute{a}jek$ \cite{hajek1968} projection:
\begin{equation}
\widetilde{\mathbf{U}}_{n}=\frac{s}{n}\sum_{i=1}^{n}E\left[  \mathbf{f}\left(
\mathbf{Y}_{1},\ldots,\mathbf{Y}_{s}\right)  \mid \mathbf{Y}_{i}\right]  .
\label{U.3}%
\end{equation}
The conditional expectations of $f\left(  \mathbf{Y}_{1},\ldots,\mathbf{Y}%
_{s}\right)  $\ given each $\mathbf{Y}_{i}$ (of the $i.i.d.$ sample) are now
$i.i.d.$, permitting applications of conventional asymptotic techniques
\cite{Kowalski2007}$.$ As shown below, the U-statistic and its projection have
the same asymptotic distribution.

\textbf{Theorem 1. }\ Let
\begin{equation}
\widetilde{\mathbf{v}}_{1}\left(  \mathbf{Y}_{1}\right)  =E\left[
\mathbf{f}\left(  \mathbf{Y}_{1},\ldots,\mathbf{Y}_{s}\right)  \mid
\mathbf{Y}_{1}\right]  -\boldsymbol{\theta},\text{ }\mathbf{e}_{n}=\sqrt
{n}\left(  \mathbf{U}_{n}-\widetilde{\mathbf{U}}_{n}\right)  ,\text{
}\mathbf{\Sigma}_{v}=Var\left[  \widetilde{\mathbf{v}}_{1}\left(
\mathbf{Y}_{1}\right)  \right]  . \label{U.4n}%
\end{equation}
Under mild regularity conditions, $\mathbf{e}_{n}\rightarrow_{p}\mathbf{0}$
and thus,

(i) $\mathbf{U}_{n}$ is consistent, i.e., $\mathbf{U}_{n}\rightarrow
_{p}\boldsymbol{\theta}.$

(ii) $\mathbf{U}_{n}$ is asymptotically (multivariate) normal:
\begin{equation}
\sqrt{n}\left(  \mathbf{U}_{n}-\boldsymbol{\theta}\right)  \rightarrow
_{d}N\left(  \mathbf{0},\mathbf{\Sigma}_{U}=s^{2}\mathbf{\Sigma}_{v}\right)  ,
\label{U.5}%
\end{equation}
where $\rightarrow_{p\text{ }(d)}$ denotes convergence in probability
(distribution). \ 

\subsubsection{U-statistics-based Generalized Estimating Equations}

For notational brevity, here we focus on between-subject, or pairwise,
attributes where the number of input vector $s=2$. Extensions to $s>2$ are
straightforward. \ 

To tackle the interlocking dependencies among pairwise outcomes for the
inference of $\boldsymbol{\beta}$, we define a class of U-statistics-based
Generalized Estimating Equations (UGEE) \cite{Kowalski2007}
\begin{align}
\mathbf{U}_{n}\left(  \boldsymbol{\beta}\right)   &  =\sum_{\mathbf{i\in}%
C_{2}^{n}}\mathbf{U}_{n,\mathbf{i}}\left(  \boldsymbol{\beta}\right)
=\sum_{\mathbf{i\in}C_{2}^{n}}\mathbf{D}_{\mathbf{i}}^{\top}V_{\mathbf{i}%
}^{-1}S_{\mathbf{i}}\left(  \boldsymbol{\beta}\right)  =\mathbf{0,}\text{
}\mathbf{i}=\left(  i_{1},i_{2}\right)  \in C_{2}^{n},\label{eqn140}\\
S_{\mathbf{i}}\left(  \boldsymbol{\beta}\right)   &  =f_{\mathbf{i}%
}-h_{\mathbf{i}}\left(  \mathbf{X}_{\mathbf{i}};\boldsymbol{\beta}\right)
,\text{ }\mathbf{D}_{\mathbf{i}}=\frac{\partial}{\partial \boldsymbol{\beta
}^{\top}}h_{\mathbf{i}}(\mathbf{X}_{\mathbf{i}};\boldsymbol{\beta}),\text{
}V_{\mathbf{i}}=Var\left(  f_{\mathbf{i}}\mid \mathbf{X}_{\mathbf{i}}\right)
.\nonumber
\end{align}
In practice, $V_{\mathbf{i}}$ is unknown and substituted by a working
variance. Estimators $\widehat{\boldsymbol{\beta}}_{\text{ugee}}$ are obtained
by solving (\ref{eqn140}) numerically such as with the Newton-Raphson method.
Although similar in appearance to GEE\  \cite{Tang2011}$,$ UGEE is not a sum of
independent variables. But $\widehat{\boldsymbol{\beta}}_{\text{ugee}}$ is
consistent and asymptotically normal, which is a direct result from Theorem 1.

\textbf{Theorem 2. }\ Let
\begin{equation}
\widetilde{\mathbf{v}}_{i_{1}}=2E\left(  \mathbf{U}_{n,\mathbf{i}}%
\mid \mathbf{Y}_{i_{1}},\mathbf{X}_{i_{1}}\right)  ,\text{ }\mathbf{B}=E\left(
\mathbf{D}_{\mathbf{i}}^{\top}V_{\mathbf{i}}^{-1}\mathbf{D}_{\mathbf{i}%
}\right)  ,\text{ }\mathbf{\Sigma}_{U}=Var\left(  \widetilde{\mathbf{v}%
}_{i_{1}}\right)  ,\text{ }\mathbf{\Sigma}_{\boldsymbol{\beta}}^{\text{ugee}%
}=\mathbf{B}^{-1}\mathbf{\Sigma}_{U}\mathbf{B}^{-1}. \label{eqn.thm1}%
\end{equation}
Under mild regularity conditions, $\widehat{\boldsymbol{\beta}}_{\text{ugee}}$
is a consistent and asymptotically normal (CAN) estimator of
$\boldsymbol{\beta}$ in (\ref{eqn 2.15}):
\[
\sqrt{n}\left(  \widehat{\boldsymbol{\beta}}_{\text{ugee}}-\boldsymbol{\beta
}\right)  \rightarrow_{d}N\left(  \mathbf{0},\mathbf{\Sigma}%
_{\boldsymbol{\beta}}^{\text{ugee}}\right)  .
\]
A consistent estimator of $\mathbf{\Sigma}_{\boldsymbol{\beta}}^{\text{ugee}}$
can be obtained by substituting consistent estimators of $\boldsymbol{\beta}$
and moment estimators of the respective quantities in (\ref{eqn.thm1}).
Conforming to the appealing features of its within-subject counterpart GEE,
UGEE also yields valid inference without explicitly delineating the
potentially more complex correlation structures.

Under suitable regularity conditions, the \textquotedblleft
sandwich\textquotedblright \ estimators from the GEE are not only consistent
and asymptotically normal but also achieve the semiparametric efficiency bound
\cite{Tsaitis2006} for the semiparametric GLM. This semiparametric efficiency
allows a sensitive signal-detection in practice while simultaneously
harmonizing robustness to model misspecification.

Our goal is to study whether the UGEE estimator for between-subject attributes
also attain the semiparametric efficiency bound for the semiparametric FRM (in
addition to CAN). We start with essential concepts and models for pairs.

\section{Asymptotic Linearity and Influence Function \label{New.secn1}}

To study the semiparametric efficiency for between-subject attributes, we
first need to extend concepts of within to between-subject attributes, such as
asymptotic linearity and influence functions \cite{Bickle1998}$.$ Denote the
within-subject attributes by $\mathbf{Z}_{1}\mathbf{,...,Z}_{n}\sim
^{i.i.d.}\left \{  p\left(  \mathbf{Z}_{i};\boldsymbol{\theta}\right)  ;\text{
}\boldsymbol{\theta}\in \Omega \right \}  $, where $p\left(  \mathbf{Z}%
_{i};\boldsymbol{\theta}\right)  $ is a probability density or distribution
function characterized by parameter $\boldsymbol{\theta}$. We assume
$\boldsymbol{\theta}=\left(  \boldsymbol{\beta}^{\top},\eta \right)  ^{\top}$
(i.e., $\boldsymbol{\beta}$ and $\eta$ are variationally independent with no
overlapping components), where $\boldsymbol{\beta}$ is a $q\times1$ vector of
parameters of interest and $\eta$ is the nuisance parameter. The only
component that differentiates parametric from semiparametric models is the
dimension of $\eta$; a finite-dimensional vector $\boldsymbol{\eta}$ yields
the parametric while an infinite-dimensional nuisance parameter, denote by
$\eta(\cdot),$ leads to the semiparametric model \cite{Tsaitis2006}$.$ \ 

In the literature ofclassical within-subject attributes, $\widehat
{\boldsymbol{\beta}}$ is asymptotically linear (AL) if there exists an
expansion $n^{1/2}\left(  \widehat{\boldsymbol{\beta}}-\boldsymbol{\beta}%
_{0}\right)  =n^{-1/2}\sum_{i=1}^{n}\boldsymbol{\phi}\mathbf{(Z}_{i}%
\mathbf{;}\boldsymbol{\theta}_{0}\mathbf{)}+o_{p}\left(  1\right)  ,$ where
$\boldsymbol{\phi}\mathbf{(Z}_{i}\mathbf{;}\boldsymbol{\theta}_{0}\mathbf{)}$
is termed the influence function (I.F.) for the $i$-th observation at
$\boldsymbol{\theta}_{0}$ (the truth). This I.F. has mean zero and finite and
nonsingular $E\left(  \boldsymbol{\phi \phi}^{\top}\right)  $, its name
reflects the influence of an observation unit on an estimator
\cite{ichimura2015}$.$\ The asymptotic normality is readily derived from this
expansion with CLT, hence, $n^{1/2}\left(  \widehat{\boldsymbol{\beta}%
}-\boldsymbol{\beta}_{0}\right)  \rightarrow_{d}N\left(  \mathbf{0},E\left(
\boldsymbol{\phi \phi^{\top}}\right)  \right)  ,$ the asymptotic variance of
$\widehat{\boldsymbol{\beta}}$ is determined by its I.F., or $\boldsymbol{\phi
}\left(  \mathbf{Z}_{i}\right)  $ defines the efficiency of $\widehat
{\boldsymbol{\beta}}$ \cite{Tsaitis2006}$.$

Those within-subject $\mathbf{Z}_{i}$'s induce a sequence of $i.d.$
(identically distributed but not necessarily independent) random vectors for
pairs, $\mathbf{Z}_{\mathbf{i}}=\left(  \mathbf{Z}_{i_{1}}^{\top}%
,\mathbf{Z}_{i_{2}}^{\top}\right)  ^{\top}\sim \left \{  p\left(  \mathbf{Z}%
_{\mathbf{i}};\boldsymbol{\theta}\right)  ;\text{ }\boldsymbol{\theta}%
\in \Omega \right \}  $. We consider two classes of models for $\mathbf{Z}%
_{\mathbf{i}}$, each with associated estimators and I.F.s.

\subsection{Non-overlap Model Class 1}

To avoid dependencies, we first consider a subset of $i.i.d.$ pairs
$\mathbf{Z}_{\mathbf{i}_{j}}$, $1\leq j\leq \lfloor \frac{n}{2}\rfloor=m$, where
$\lfloor \cdot \rfloor$ is the floor function. Namely, we reorganize the data
into independent non-overlapping pairs. Although this reorganization is not
unique, we choose one without loss of generality. For example, when $n=4$, we
can choose $\{ \mathbf{Z}_{i_{1}}\mathbf{,Z}_{i_{2}}\}$ and $\{ \mathbf{Z}%
_{i_{3}}\mathbf{,Z}_{i_{4}}\}$ to form two independent pairs $i_{1}=$ $\left(
i_{1},i_{2}\right)  $ and $i_{2}=$ $\left(  i_{3},i_{4}\right)  $. This
removes the hurdle of dependencies originated from overlapping pairs,
permitting definitions in parallel with the classical setting.

\textbf{Definition.} $\widetilde{\boldsymbol{\beta}}$ is an asymptotically
linear (AL) estimator of between-subject attributes for the non-overlap model
class 1 if it belongs to
\begin{equation}
\Omega_{1}^{\boldsymbol{\beta}}=\left \{  \widetilde{\boldsymbol{\beta}%
}(\mathbf{Z}_{\mathbf{i}_{j}}):\sqrt{m}\left(  \widetilde{\boldsymbol{\beta}%
}-\boldsymbol{\beta}_{0}\right)  =\sqrt{m}\frac{1}{m}\sum_{j=1}^{m}%
\boldsymbol{\psi}\left(  \mathbf{Z}_{\mathbf{i}_{j}};\boldsymbol{\theta}%
_{0}\right)  +\mathbf{o}_{p}(1)\right \}  , \label{model1}%
\end{equation}
where $\boldsymbol{\psi}^{q\times1}\left(  \mathbf{Z}_{\mathbf{i}%
};\boldsymbol{\theta}_{0}\right)  $ is a measurable function with mean zero,
finite and nonsingular $E\left(  \boldsymbol{\psi \psi}^{\top}\right)  $,
termed the influence function 1 for the $i$-th pair at the truth. The set of
all influence functions 1is denoted by $\Gamma_{1}^{I.F.}.$

Under mild regularity conditions, CLT yields
\[
\sqrt{m}\left(  \widetilde{\boldsymbol{\beta}}-\boldsymbol{\beta}_{0}\right)
\rightarrow_{d}N\left(  \mathbf{0},\mathbf{\Sigma}_{1}\right)  ,\mathbf{\Sigma
}_{1}=E\left[  \boldsymbol{\psi}\left(  \mathbf{Z}_{\mathbf{i}_{j}}\right)
\boldsymbol{\psi}^{\top}\left(  \mathbf{Z}_{\mathbf{i}_{j}}\right)  \right]
.
\]
Thus, the asymptotic variance for $\widetilde{\boldsymbol{\beta}}$ $\in
\Omega_{1}^{\boldsymbol{\beta}}$ is determined by its I.F. $\boldsymbol{\psi
}\left(  \mathbf{Z}_{\mathbf{i}_{j}}\right)  $, and the efficient estimator in
$\Omega_{1}^{\boldsymbol{\beta}}$ is the one with minimum variance.

In practice, we do not fit data with between-subject attributes using model
class 1 as they only deploy part of the data. But this conceptual model will
help us pinpoint the efficient estimator for the FRM as we now introduce.

\subsection{Enumerated Model Class 2}

If making the inference based on all possible pairs $\mathbf{Z}_{\mathbf{i}}$,
$\mathbf{i}=\left(  i_{1},i_{2}\right)  \in C_{2}^{n}$, including those with
overlapping subjects, we reimpose the dependencies and form a class of
enumerated model 2. The FRM in (\ref{eqn 2.15})\ that engage all possible
pairs is an example of this class.

\textbf{Definition.} $\widehat{\boldsymbol{\beta}}$ is an AL estimator of
between-subject attributes for the enumerated model 2 if it belongs to%
\begin{equation}
\Omega_{2}^{\boldsymbol{\beta}}=\left \{  \widehat{\boldsymbol{\beta}%
}(\mathbf{Z}_{\mathbf{i}}):n^{1/2}\left(  \widehat{\boldsymbol{\beta}%
}-\boldsymbol{\beta}_{0}\right)  =\sqrt{n}\binom{n}{2}^{-1}\sum_{\mathbf{i}\in
C_{2}^{n}}\boldsymbol{\varphi}\left(  \mathbf{Z}_{\mathbf{i}}%
;\boldsymbol{\theta}_{0}\right)  +\mathbf{o}_{p}\left(  1\right)  \right \}  ,
\label{eqn.n10}%
\end{equation}
where the measurable function $\boldsymbol{\varphi}^{q\times1}\left(
\mathbf{Z}_{\mathbf{i}};\boldsymbol{\theta}_{0}\right)  $ with mean zero and
finite and nonsingular $E\left(  \boldsymbol{\varphi \varphi}^{\top}\right)  $
is defined as the enumerated influence function 2 for the $\mathbf{i}$-th
pair\ at the truth. Denote the set of all such $\boldsymbol{\varphi}\left(
\mathbf{Z}_{\mathbf{i}};\boldsymbol{\theta}_{0}\right)  $ by $\Gamma
_{2}^{I.F.}.$ It is obvious that $\widehat{\boldsymbol{\beta}}_{\text{ugee}%
}\in \Omega_{2}^{\boldsymbol{\beta}}.$

As (\ref{eqn.n10}) involves the summation of dependent $\boldsymbol{\varphi
}\left(  \mathbf{Z}_{\mathbf{i}}\right)  $, we apply (\ref{U.4n}) to obtain:
\
\begin{align}
\sqrt{n}\left(  \widehat{\boldsymbol{\beta}}-\boldsymbol{\beta}_{0}\right)
&  =\sqrt{n}\frac{1}{n}\sum \limits_{i_{1}=1}^{n}2E\left[  \boldsymbol{\varphi
}\left(  \mathbf{Z}_{\mathbf{i}};\boldsymbol{\theta}_{0}\right)
\mid \mathbf{Z}_{i_{1}}\right]  +\mathbf{o}_{p}\left(  1\right)  \rightarrow
_{d}N\left(  \mathbf{0},\mathbf{\Sigma}_{2}\right)  ,\nonumber \\
\mathbf{\Sigma}_{2}  &  =Var\left \{  2E\left[  \boldsymbol{\varphi}\left(
\mathbf{Z}_{\mathbf{i}};\boldsymbol{\theta}_{0}\right)  \mid \mathbf{Z}_{i_{1}%
}\right]  \right \}  =E\left \{  2E\left[  \boldsymbol{\varphi}\left(
\mathbf{Z}_{\mathbf{i}};\boldsymbol{\theta}_{0}\right)  \mid \mathbf{Z}_{i_{1}%
}\right]  \cdot2E\left[  \boldsymbol{\varphi}^{\top}\left(  \mathbf{Z}%
_{\mathbf{i}};\boldsymbol{\theta}_{0}\right)  \mid \mathbf{Z}_{i_{1}}\right]
\right \}  . \label{CH03.eqn3.12000}%
\end{align}
Hence, for $\widehat{\boldsymbol{\beta}}\in \Omega_{2}^{\boldsymbol{\beta}}$ of
model class 2, the asymptotic variance $\mathbf{\Sigma}_{2}$ is also
determined by its I.F.. $\mathbf{\Sigma}_{2}$ apparently differs from
$\mathbf{\Sigma}_{1}$ in model class 1, since it involves an additional step
of mapping from a function of between-subject attribute $\mathbf{Z}%
_{\mathbf{i}}$ to a function of within-subject attribute $\mathbf{Z}_{i_{1}}.$

\subsection{Relationships between I.F.s for the Two Model Classes}

The influence function is the key to studying efficiency, we now elucidate the
relationships between influence functions associated with each model class.

\textbf{Equivalence of Two Classes of AL estimators }

For AL estimators, the I.F.s for the two model classes are equivalent. Namely,
for any $\boldsymbol{\psi}\left(  \mathbf{Z}_{\mathbf{i}};\boldsymbol{\theta
}_{0}\right)  \in \Gamma_{1}^{I.F.}$ and AL\ estimator $\widetilde
{\boldsymbol{\beta}}\in \Omega_{1}^{\boldsymbol{\beta}}$, we can construct
another estimator $\widehat{\boldsymbol{\beta}}=\boldsymbol{\beta}_{0}%
+\binom{n}{2}^{-1}\sum_{\mathbf{i}\in C_{2}^{n}}\boldsymbol{\psi}\left(
\mathbf{Z}_{\mathbf{i}};\boldsymbol{\theta}_{0}\right)  .$ It is readily
checked that this new $\widehat{\boldsymbol{\beta}}\in \Omega_{2}%
^{\boldsymbol{\beta}}$ by satisfying (\ref{eqn.n10}), indicating that
$\widehat{\boldsymbol{\beta}}$ is also AL for the model class 2 and
$\boldsymbol{\psi}\left(  \mathbf{Z}_{\mathbf{i}};\boldsymbol{\theta}%
_{0}\right)  \in \Gamma_{2}^{I.F.}.$ \ 

Conversely, for any $\boldsymbol{\varphi}\left(  \mathbf{Z}_{\mathbf{i}%
};\boldsymbol{\theta}_{0}\right)  \in \Gamma_{2}^{I.F.}$ and corresponding
AL\ estimator $\widehat{\boldsymbol{\beta}}$ $\in \Omega_{2}^{\boldsymbol{\beta
}}$ satisfying (\ref{eqn.n10}), we define an estimator $\widetilde
{\boldsymbol{\beta}}=\boldsymbol{\beta}_{0}+m^{-1}\sum_{j=1}^{m}%
\boldsymbol{\varphi}\left(  \mathbf{Z}_{\mathbf{i}_{j}};\boldsymbol{\theta
}_{0}\right)  .$ It is again readily shown that this estimator satisfies
(\ref{model1}), indicating that $\widetilde{\boldsymbol{\beta}}\in \Omega
_{1}^{\boldsymbol{\beta}}$ and $\boldsymbol{\varphi}\left(  \mathbf{Z}%
_{\mathbf{i}};\boldsymbol{\theta}_{0}\right)  \in \Gamma_{1}^{I.F.},$ i.e.,
$\boldsymbol{\varphi}\left(  \mathbf{Z}_{\mathbf{i}};\boldsymbol{\theta}%
_{0}\right)  $ is also an I.F. for model class 1.

\textbf{Equivalence of Two Classes of Regular and AL estimators }

As in the literature, to avoid estimators with undesirable local properties
such as super-efficiency \cite{lecam1953}$,$ we restrict considerations to
regular estimators by considering a local data generating process (LDGP).
Suppose the underlying within-subject attributes are generated from
$\mathbf{Z}_{in}\sim^{i.i.d.}\left \{  p\left(  \mathbf{Z}_{in}%
;\boldsymbol{\theta}_{n}\right)  \right \}  $ for each $\boldsymbol{\theta}%
_{n}$, and $n^{1/2}\left(  \boldsymbol{\theta}_{n}-\boldsymbol{\theta}_{\ast
}\right)  $ converges to a constant where $\boldsymbol{\theta}_{\ast}$ denote
some fixed parameter. Let $\widehat{\boldsymbol{\theta}}(\mathbf{Z}%
_{\mathbf{i}n})$ denote an estimator of $\boldsymbol{\theta}_{n}$ based on the
between-subject attributes, where $\mathbf{Z}_{\mathbf{i}n}\mathbf{=(Z}%
_{i_{1}n}^{\top}\mathbf{,Z}_{i_{2}n}^{\top}\mathbf{)}^{\top}$. Then
$\widehat{\boldsymbol{\theta}}(\mathbf{Z}_{\mathbf{i}n})$ is regular if, for
some fixed $\boldsymbol{\theta}_{\ast}$, the limiting distribution of
$n^{1/2}\left(  \widehat{\boldsymbol{\theta}}(\mathbf{Z}_{\mathbf{i}%
n})-\boldsymbol{\theta}_{n}\right)  $ does not depend on the LDGP (or
$\boldsymbol{\theta}_{n}$). In what follows, we focus on regular
and\ asymptotically linear (RAL) estimators unless stated otherwise. The
theorem below further declares the equivalence between the two classes of
I.F.s for Regular AL estimators.

\textbf{Theorem 3.} For RAL estimators of between-subject attributes, the
I.F.s in the set $\Gamma_{1}^{I.F.}$ for model class 1 are equivalent to I.F.s
in $\Gamma_{2}^{I.F.}$ for model class 2, i.e., $\Gamma_{1}^{I.F.}=\Gamma
_{2}^{I.F.}.$

Although we aim to find the efficient I.F. for the FRM in model class 2, it is
difficult to directly work with model class 2 due to the added complexity in
computing asymptotic variance through $H\acute{a}jek$ projection. Accordingly,
Theorem 3 can allow us to achieve the goal by virtue of the simplicity of
model class 1.

\section{The Hilbert Space and Projection\label{Ef.sec4}}

In this section, we start with a brief review of Hilbert space
\cite{walter1987} and its application to the within-subject attributes and
then extend it to their between-subject counterparts. More details can be
found in the Supplement S1.

\subsection{Within-subject Attributes}

Let $(L,A,P)$ be a probability space (where $L$ is the sample space, $A$ is
the $\sigma$-algebra, and $P$ is the probability measure). Consider a
$q$-dimensional measurable function $\mathbf{Z}:$ $L$ $\rightarrow$ $%
%TCIMACRO{\U{211d} }%
%BeginExpansion
\mathbb{R}
%EndExpansion
^{q}$. Suppose we observe $i.i.d.$ within-subject attributes $\mathbf{Z}%
_{1}\mathbf{,...,Z}_{n}$, where $\mathbf{Z}_{i}$ is the random vector for
subject $i$. We denote by $\mathcal{H}_{w}$ the Hilbert space consisting of
all $q$-dimensional functions of $\mathbf{Z}_{i}$, $\mathbf{h}:L\rightarrow%
%TCIMACRO{\U{211d} }%
%BeginExpansion
\mathbb{R}
%EndExpansion
^{q},$ that are measurable with mean zero and finite second-order moments.
$\mathcal{H}_{w}$ is associated with an inner product, which also induces a
norm (we emphasize quantities of within-subject attributes with a subscript
$w$) :%
\begin{equation}
\left \langle \mathbf{h}_{1}\left(  \mathbf{Z}_{i}\right)  ,\mathbf{h}%
_{2}\left(  \mathbf{Z}_{i}\right)  \right \rangle _{w}=E\left[  \mathbf{h}%
_{1}^{\top}\left(  \mathbf{Z}_{i}\right)  \mathbf{h}_{2}\left(  \mathbf{Z}%
_{i}\right)  \right]  ,\text{ }\left \Vert \mathbf{h}\left(  \mathbf{Z}%
_{i}\right)  \right \Vert _{w}=\left \langle \mathbf{h},\mathbf{h}\right \rangle
_{w}^{1/2}=E^{1/2}\left[  \mathbf{h}^{\top}\left(  \mathbf{Z}_{i}\right)
\mathbf{h}\left(  \mathbf{Z}_{i}\right)  \right]  . \label{eqn.w.01}%
\end{equation}

Let $\mathbf{v}\left(  \mathbf{Z}_{i}\right)  =\left(  v_{1}\left(
\mathbf{Z}_{i}\right)  ,...,v_{r}\left(  \mathbf{Z}_{i}\right)  \right)
^{\top}$ be an $r$-dimensional random function with $E\left[  \mathbf{v}%
\left(  \mathbf{Z}_{i}\right)  \right]  =0$ and $\left \langle \mathbf{v}%
,\mathbf{v}\right \rangle _{w}<\infty$. For the linear subspace spanned by
$\mathbf{v}\left(  \mathbf{Z}_{i}\right)  $: $\ $%
\[
\mathcal{U}_{w}=\{ \mathbf{Bv}\left(  \mathbf{Z}_{i}\right)  ;\text{ for an
arbitrary matrix }\mathbf{B}^{q\times r\text{ }}\text{ of real numbers}\},
\]
by the closest point theorem \cite{sehgal1987}$,$ the projection of
$\mathbf{h}^{q\times1\text{ }}\left(  \mathbf{Z}_{i}\right)  \in
\mathcal{H}_{w}$ onto $\mathcal{U}_{w}$ is unique, denote by\
\begin{equation}
\Pi_{w}\left \{  \mathbf{h}\left(  \mathbf{Z}_{i}\right)  \mid \mathcal{U}%
_{w}\right \}  =E\left[  \mathbf{h}\left(  \mathbf{Z}_{i}\right)
\mathbf{v}\left(  \mathbf{Z}_{i}\right)  ^{\top}\right]  E^{-1}\left[
\mathbf{v}\left(  \mathbf{Z}_{i}\right)  \mathbf{v}\left(  \mathbf{Z}%
_{i}\right)  ^{\top}\right]  \mathbf{v}\left(  \mathbf{Z}_{i}\right)  .
\label{eqn.w.02}%
\end{equation}

\subsection{Between-subject Attributes \label{Ef.sec4.00}}

For the induced pairwise observations $\mathbf{Z}_{\mathbf{i}}=\mathbf{(Z}%
_{i_{1}}^{\top}\mathbf{,Z}_{i_{2}}^{\top}\mathbf{)}^{\top}$, we consider the
Hilbert space $\mathcal{H}_{b}$ (with a subscript $b$ reflecting
between-subject attributes) of all $q$-dimensional measurable and symmetric
functions $\mathbf{h}\left(  \mathbf{Z}_{\mathbf{i}}\right)  =\mathbf{h}%
\left(  \mathbf{Z}_{i_{1}},\mathbf{Z}_{i_{2}}\right)  $ with $E\left[
\mathbf{h}\left(  \mathbf{Z}_{\mathbf{i}}\right)  \right]  =\mathbf{0}$ and
finite $E\left[  \mathbf{h}\left(  \mathbf{Z}_{\mathbf{i}}\right)
\mathbf{h}^{\top}\left(  \mathbf{Z}_{\mathbf{i}}\right)  \right]  $. We
consider two inner products and norms for $\mathcal{H}_{b}$. \ 

\textbf{Definition.} The non-overlap inner product b1 and associated norm b1
are defined as%
\begin{align}
\left \langle \mathbf{h}_{1}\left(  \mathbf{Z}_{\mathbf{i}}\right)
,\mathbf{h}_{2}\left(  \mathbf{Z}_{\mathbf{i}}\right)  \right \rangle _{b1}  &
=E\left[  \mathbf{h}_{1}^{\top}\left(  \mathbf{Z}_{\mathbf{i}}\right)
\mathbf{h}_{2}\left(  \mathbf{Z}_{\mathbf{i}}\right)  \right]  ,\text{
}\label{eqn.b.01}\\
\left \Vert \mathbf{h}\left(  \mathbf{Z}_{\mathbf{i}}\right)  \right \Vert
_{b1}  &  =\left \langle \mathbf{h}\left(  \mathbf{Z}_{\mathbf{i}}\right)
,\mathbf{h}\left(  \mathbf{Z}_{\mathbf{i}}\right)  \right \rangle _{b1}%
^{1/2}=E^{1/2}\left[  \mathbf{h}^{\top}\left(  \mathbf{Z}_{\mathbf{i}}\right)
\mathbf{h}\left(  \mathbf{Z}_{\mathbf{i}}\right)  \right]  .\nonumber
\end{align}
For the linear span of $\mathbf{v}\left(  \mathbf{Z}_{\mathbf{i}}\right)
=\left(  v_{1}\left(  \mathbf{Z}_{\mathbf{i}}\right)  ,...,v_{r}\left(
\mathbf{Z}_{\mathbf{i}}\right)  \right)  ^{\top}$ (as a function of
$\mathbf{Z}_{\mathbf{i}}$ for the $\mathbf{i}$-th pair):%
\[
\mathcal{U}_{b1}=\{ \mathbf{Bv(\mathbf{Z}_{\mathbf{i}})};\text{ for an
arbitrary matrix }\mathbf{B}^{q\times r\text{ }}\text{ of real numbers}\},
\]
the projection of $\mathbf{h}^{q\times1}\left(  \mathbf{Z}_{\mathbf{i}%
}\right)  \in \mathcal{H}_{b}$ onto $\mathcal{U}_{b1}$ is:
\begin{equation}
\Pi_{b1}\left \{  \mathbf{h}\left(  \mathbf{Z}_{\mathbf{i}}\right)
\mid \mathcal{U}_{b1}\right \}  =E\left[  \mathbf{h}\left(  \mathbf{Z}%
_{\mathbf{i}}\right)  \mathbf{v}\left(  \mathbf{Z}_{\mathbf{i}}\right)
^{\top}\right]  E^{-1}\left[  \mathbf{v}\left(  \mathbf{Z}_{\mathbf{i}%
}\right)  \mathbf{v}\left(  \mathbf{Z}_{\mathbf{i}}\right)  ^{\top}\right]
\mathbf{v}\left(  \mathbf{Z}_{\mathbf{i}}\right)  . \label{eqn.b.02}%
\end{equation}
It follows from the Pythagorean triangle inequality that
\begin{equation}
\left \Vert \mathbf{h}\left(  \mathbf{Z}_{\mathbf{i}}\right)  \right \Vert
_{b1}^{2}=\left \Vert \Pi_{b1}\left \{  \mathbf{h}\mid \mathcal{U}_{b1}\right \}
\right \Vert _{b1}^{2}+\left \Vert \mathbf{\mathbf{h}}-\Pi_{b1}\left \{
\mathbf{h}\mid \mathcal{U}_{b1}\right \}  \right \Vert _{b1}^{2}\geq \left \Vert
\Pi_{b1}\left \{  \mathbf{h}\left(  \mathbf{Z}_{\mathbf{i}}\right)
\mid \mathcal{U}_{b1}\right \}  \right \Vert _{b1}^{2}, \label{normb1}%
\end{equation}
i.e., the norm b1 of any element $\mathbf{h}\left(  \mathbf{Z}_{\mathbf{i}%
}\right)  $ is larger than or equal to that of its projection onto the
subspace $\mathcal{U}_{b1}$.

Under the $q$-replicating linear spaces \cite{Tsaitis2006} that we consider,
the multivariate Pythagoras holds:\ the orthogonality between $\mathbf{h}%
\left(  \mathbf{Z}_{\mathbf{i}}\right)  $ and $\mathcal{U}_{b1}$ is equivalent
to the uncorrelatedness between $\mathbf{h}\left(  \mathbf{Z}_{\mathbf{i}%
}\right)  $ and $\mathbf{v}\left(  \mathbf{Z}_{\mathbf{i}}\right)  $ (i.e.,
$E\left[  \mathbf{h}^{\top}\left(  \mathbf{Z}_{\mathbf{i}}\right)
\mathbf{v}\left(  \mathbf{Z}_{\mathbf{i}}\right)  \right]  =0$ implies
$E\left[  \mathbf{h}\left(  \mathbf{Z}_{\mathbf{i}}\right)  \mathbf{v}^{\top
}\left(  \mathbf{Z}_{\mathbf{i}}\right)  \right]  =0$). Thus (\ref{normb1})
shows that the variance (matrix) of the element $\mathbf{h}\left(
\mathbf{Z}_{\mathbf{i}}\right)  \in \mathcal{H}_{b}$ also satisfies%
\begin{equation}
Var\left[  \mathbf{\mathbf{h}}\left(  \mathbf{Z}_{\mathbf{i}}\right)  \right]
=Var\left[  \Pi_{b1}\left \{  \mathbf{h}\mid \mathcal{U}_{b1}\right \}  \right]
+Var\left[  \mathbf{\mathbf{h}}-\Pi_{b1}\left \{  \mathbf{h}\mid \mathcal{U}%
_{b1}\right \}  \right]  \geq Var\left[  \Pi_{b1}\left \{  \mathbf{h}\left(
\mathbf{Z}_{\mathbf{i}}\right)  \mid \mathcal{U}_{b1}\right \}  \right]  .
\label{Var1}%
\end{equation}
Hence, the variance of any element $\mathbf{h}\left(  \mathbf{Z}_{\mathbf{i}%
}\right)  $ is larger than or equal to its projection $\Pi_{b1}\left \{
\mathbf{h}\mid \mathcal{U}_{b1}\right \}  $ onto a subspace, i.e., their
difference is non-negative definite. This will inspire the construction of the
efficient estimator using the projection later.

As the norm b1 for $\mathcal{H}_{b}$ does not yield the asymptotic variance
for the UGEE estimator, we now introduce another inner product motivated by
the form of asymptotic variance for the enumerated model class 2.

\textbf{Definition.} The enumerated inner product 2 and norm b2 are defined as%
\begin{align}
\left \langle \mathbf{h}_{1}\left(  \mathbf{Z}_{\mathbf{i}}\right)
,\mathbf{h}_{2}\left(  \mathbf{Z}_{\mathbf{i}}\right)  \right \rangle _{b2}  &
=E\left \{  2E\left[  \mathbf{h}_{1}^{\top}\left(  \mathbf{Z}_{\mathbf{i}%
}\right)  \mid \mathbf{Z}_{i_{1}}\right]  \cdot2E\left[  \mathbf{h}_{2}\left(
\mathbf{Z}_{\mathbf{i}}\right)  \mid \mathbf{Z}_{i_{1}}\right]  \right \}
,\label{eqn.b.03}\\
\left \Vert \mathbf{h}\left(  \mathbf{Z}_{\mathbf{i}}\right)  \right \Vert
_{b2}  &  =\left \langle \mathbf{h}\left(  \mathbf{Z}_{\mathbf{i}}\right)
,\mathbf{h}\left(  \mathbf{Z}_{\mathbf{i}}\right)  \right \rangle _{b2}%
^{1/2}=E^{1/2}\left \{  2E\left[  \mathbf{h}^{\top}\left(  \mathbf{Z}%
_{\mathbf{i}}\right)  \mid \mathbf{Z}_{i_{1}}\right]  \cdot2E\left[
\mathbf{h}\left(  \mathbf{Z}_{\mathbf{i}}\right)  \mid \mathbf{Z}_{i_{1}%
}\right]  \right \}  .\nonumber
\end{align}

\textbf{Definition.} Define a projection mapping \cite{luenberger1997}
$\mathcal{M}$: $\mathcal{H}_{b}\rightarrow \mathcal{H}_{w}$, referred to as the
U-statistics, or $H\acute{a}jek$, projection, such that for $\mathbf{h}\left(
\mathbf{Z}_{\mathbf{i}}\right)  \in \mathcal{H}_{b}$,
\begin{equation}
\mathcal{M}\left[  \mathbf{h}\left(  \mathbf{Z}_{\mathbf{i}}\right)  \right]
=2E\left[  \mathbf{h}\left(  \mathbf{Z}_{\mathbf{i}}\right)  \mid
\mathbf{Z}_{i_{1}}\right]  \in \mathcal{H}_{w}. \label{U-mapping}%
\end{equation}

Now consider a linear subspace of $\mathcal{H}_{w}$ spanned by $\mathcal{M}%
\left[  \mathbf{B}^{q\times r\text{ }}\mathbf{v}\left(  \mathbf{Z}%
_{\mathbf{i}}\right)  \right]  =\mathbf{B}E\left[  \mathbf{v}\left(
\mathbf{Z}_{\mathbf{i}}\right)  \mid \mathbf{Z}_{i_{1}}\right]  $,%
\[
\mathcal{U}_{b2}=\mathcal{M}\left(  \mathcal{U}_{b1}\right)  =\{
\mathbf{B}E\left[  \mathbf{v}\left(  \mathbf{Z}_{\mathbf{i}}\right)
\mid \mathbf{Z}_{i_{1}}\right]  ;\text{ for an arbitrary matrix }%
\mathbf{B}^{q\times r\text{ }}\text{ of real numbers}\}.
\]
Projecting any $\mathbf{h}\left(  \mathbf{Z}_{\mathbf{i}}\right)
\in \mathcal{H}_{b}$ onto $\mathcal{U}_{b2}$ involves two steps:\ we first map
$\mathbf{h}\left(  \mathbf{Z}_{\mathbf{i}}\right)  $ to $\mathcal{M}\left[
\mathbf{h}\left(  \mathbf{Z}_{\mathbf{i}}\right)  \right]  \in \mathcal{H}_{w}$
and then project it onto $\mathcal{U}_{b2}$ with the projection theorem for
within-subject attributes in (\ref{eqn.w.02}), i.e.:
\begin{equation}
\Pi_{b2}\left \{  \mathbf{h}\left(  \mathbf{Z}_{\mathbf{i}}\right)
\mid \mathcal{U}_{b2}\right \}  =\Pi_{w}\left \{  \mathcal{M}\left[
\mathbf{h}\left(  \mathbf{Z}_{\mathbf{i}}\right)  \right]  \mid \mathcal{U}%
_{b2}\right \}  . \label{normb222}%
\end{equation}
Similarly, the norm b2 of any element $\mathbf{h}\left(  \mathbf{Z}%
_{\mathbf{i}}\right)  $is larger than or equal to that of its projection onto
$\mathcal{U}_{b2}$, which by (\ref{normb222}), equals the squared norm of the
mapped element $\mathcal{M}\left[  \mathbf{h}\left(  \mathbf{Z}_{\mathbf{i}%
}\right)  \right]  ,$ i.e.,
\begin{equation}
\left \Vert \mathbf{h}\left(  \mathbf{Z}_{\mathbf{i}}\right)  \right \Vert
_{b2}^{2}\geq \left \Vert \Pi_{b2}\left \{  \mathbf{h}\left(  \mathbf{Z}%
_{\mathbf{i}}\right)  \mid \mathcal{U}_{b2}\right \}  \right \Vert _{b2}%
^{2}=\left \Vert \Pi_{w}\left \{  \mathcal{M}\left[  \mathbf{h}\left(
\mathbf{Z}_{\mathbf{i}}\right)  \right]  \mid \mathcal{U}_{b2}\right \}
\right \Vert _{w}^{2}. \label{normb22}%
\end{equation}

Accordingly, the variance of any $\mathcal{M}\left[  \mathbf{h}\left(
\mathbf{Z}_{\mathbf{i}}\right)  \right]  $ is larger than or equal to its
projection $\Pi_{w}\left \{  \mathcal{M}\left[  \mathbf{h}\left(
\mathbf{Z}_{\mathbf{i}}\right)  \right]  \mid \mathcal{U}_{b2}\right \}  $:%
\begin{equation}
Var\left \{  \mathcal{M}\left[  \mathbf{h}\left(  \mathbf{Z}_{\mathbf{i}%
}\right)  \right]  \right \}  \geq Var\left[  \Pi_{w}\left \{  \mathcal{M}%
\left[  \mathbf{h}\left(  \mathbf{Z}_{\mathbf{i}}\right)  \right]
\mid \mathcal{U}_{b2}\right \}  \right]  =Var\left[  \Pi_{b2}\left \{
\mathbf{h}\left(  \mathbf{Z}_{\mathbf{i}}\right)  \mid \mathcal{U}%
_{b2}\right \}  \right]  . \label{Var2}%
\end{equation}
The above links norm b2 with the asymptotic variance of the UGEE estimator. We
now define equivalence classes within each norm and discuss the relationship
between the two. \ 

\textbf{Definition.} For a given norm, any two functions $\mathbf{h}%
_{1}\left(  \mathbf{Z}_{\mathbf{i}}\right)  $ and $\mathbf{h}_{2}\left(
\mathbf{Z}_{\mathbf{i}}\right)  $ are equivalent if the norm of their
difference is zero. The equivalence class of $\mathbf{h}\left(  \mathbf{Z}%
_{\mathbf{i}}\right)  $ under norm b1 includes all $q$-dimensional measurable
functions $\mathbf{g}\left(  \mathbf{Z}_{\mathbf{i}}\right)  \in
\mathcal{H}_{b}$ that equal $\mathbf{h}\left(  \mathbf{Z}_{\mathbf{i}}\right)
$ almost surely (a.s.), denote by:%
\[
\Gamma_{b1}^{\mathbf{h}}=\left \{  \mathbf{g}\left(  \mathbf{Z}_{\mathbf{i}%
}\right)  \in \mathcal{H}_{b}:\mathbf{g}\left(  \mathbf{Z}_{\mathbf{i}}\right)
=\mathbf{h}\left(  \mathbf{Z}_{\mathbf{i}}\right)  \text{ a.s.}\right \}  .
\]
The equivalence class under norm b2 contains all functions $\mathbf{g}\left(
\mathbf{Z}_{\mathbf{i}}\right)  \in \mathcal{H}_{b}$ whose U-statistics
projection mapping are equal to that of $\mathbf{h}\left(  \mathbf{Z}%
_{\mathbf{i}}\right)  $ a.s.:
\[
\Gamma_{b2}^{\mathbf{h}}=\left \{  \mathbf{g}\left(  \mathbf{Z}_{\mathbf{i}%
}\right)  \in \mathcal{H}_{b}:\mathcal{M}\left[  \mathbf{g}\left(
\mathbf{Z}_{\mathbf{i}}\right)  \right]  =\mathcal{M}\left[  \mathbf{h}\left(
\mathbf{Z}_{\mathbf{i}}\right)  \right]  \text{ a.s.}\right \}  .
\]

The projections onto subspaces,\ $\Pi_{b1}\left \{  \mathbf{h}\left(
\mathbf{Z}_{\mathbf{i}}\right)  \mid \mathcal{U}_{b1}\right \}  $ and $\Pi
_{b2}\left \{  \mathbf{h}\left(  \mathbf{Z}_{\mathbf{i}}\right)  \mid
\mathcal{U}_{b2}\right \}  ,$ are\ unique up to their respective equivalence
classes $\Gamma_{b1}^{\mathbf{h}}$ and $\Gamma_{b2}^{\mathbf{h}}$. Since all
estimators in the same equivalence class deliver the same asymptotic variance
(or efficiency) under the respective norm, it suffices to find one of them.
Since the projection mapping $\mathcal{M}$ is many-to-one, i.e., different
elements in $\mathcal{H}_{b}$ can be mapped to the same element in
$\mathcal{H}_{w}$, the origin\ of $\mathcal{H}_{b}$ under inner product 2 is
not the equivalence class of $\mathbf{h}\left(  \mathbf{Z}_{\mathbf{i}%
}\right)  $ with $\mathbf{h}\left(  \mathbf{Z}_{\mathbf{i}}\right)  =0$ a.s.,
but a larger one consisting of functions $\mathbf{h}\left(  \mathbf{Z}%
_{\mathbf{i}}\right)  $ such that $\mathcal{M}\left[  \mathbf{h}\left(
\mathbf{Z}_{\mathbf{i}}\right)  \right]  =0$ a.s. (see Supplement for an
example of $\mathbf{h}\left(  \mathbf{Z}_{\mathbf{i}}\right)  \neq0$,
but\ $\mathcal{M}\left[  \mathbf{h}\left(  \mathbf{Z}_{\mathbf{i}}\right)
\right]  =0$\ a.s.).

Akin to the classical theory for within-subject attributes, the I.F.
$\boldsymbol{\psi}\left(  \mathbf{Z}_{\mathbf{i}};\boldsymbol{\theta}%
_{0}\right)  $ for model class 1 is an element in $\mathcal{H}_{b},$ whose
norm b1 is always larger than or equal to its projection onto a subspace
$\mathcal{U}_{b1}$, hence, this projection $\Pi_{b1}\left \{  \boldsymbol{\psi
}\left(  \mathbf{Z}_{\mathbf{i}};\boldsymbol{\theta}_{0}\right)
\mid \mathcal{U}_{b1}\right \}  $ yields an RAL estimator with the minimum
variance within class 1. Likewise, for model class 2, the projection of an
I.F. onto $\mathcal{U}_{b2}$, $\Pi_{b2}\left \{  \boldsymbol{\varphi}\left(
\mathbf{Z}_{\mathbf{i}};\boldsymbol{\theta}_{0}\right)  \mid \mathcal{U}%
_{b2}\right \}  ,$ has the smallest norm b2 thus also yields the efficient RAL
estimator for class 2. Both $\Pi_{b1}\left \{  \boldsymbol{\psi}\left(
\mathbf{Z}_{\mathbf{i}};\boldsymbol{\theta}_{0}\right)  \mid \mathcal{U}%
_{b1}\right \}  $ and $\Pi_{b2}\left \{  \boldsymbol{\varphi}\left(
\mathbf{Z}_{\mathbf{i}};\boldsymbol{\theta}_{0}\right)  \mid \mathcal{U}%
_{b2}\right \}  $ are unique up to their respective equivalence classes.

\section{Tangent Spaces and Dual Geometric Interpretations\label{New.sec3}}

The Hilbert space repositions searching for the efficient estimator to a
geometric problem of searching for the (efficient) influence function, which
has the smallest norm. Another tool we implement is a \textquotedblleft
bridge\textquotedblright \ between parametric and semiparametric models, termed
\textquotedblleft parametric submodels\textquotedblright \  \cite{Newey1990}$.$
We extend this idea to between-subject attributes next.

\subsection{Parametric Submodels \label{Ef.sec4.0}}

The distribution of pairwise observations $\mathbf{Z}_{\mathbf{i}}=\left(
\mathbf{Z}_{i_{1}}^{\top},\mathbf{Z}_{i_{2}}^{\top}\right)  ^{\top}$ can be
characterized by $p_{\mathbf{Z}}\left(  \mathbf{Z}_{\mathbf{i}}\right)  $ that
belongs to
\begin{equation}
\mathcal{P}=\left \{  p_{\mathbf{Z}}\left(  \mathbf{Z}_{\mathbf{i}%
};\boldsymbol{\beta},\eta \left(  \cdot \right)  \right)  ;\text{ }%
\boldsymbol{\beta}\in%
%TCIMACRO{\U{211d} }%
%BeginExpansion
\mathbb{R}
%EndExpansion
^{q}\text{ and }\eta \left(  \cdot \right)  \  \text{is infinite-dimensional.}%
\right \}  \label{eqn.semi.01}%
\end{equation}
Let $p_{0}\left(  \mathbf{Z}_{\mathbf{i}};\boldsymbol{\theta}_{0}\right)
=p_{\mathbf{Z}}\left(  \mathbf{Z}_{\mathbf{i}};\boldsymbol{\beta}_{0},\eta
_{0}(\cdot)\right)  $ denote the truth, where $\boldsymbol{\beta}$ and
$\eta \left(  \cdot \right)  $ are variationally independent as indicated
previously. The infinite-dimensional nuisance parameter $\eta \left(
\cdot \right)  \ $makes $\mathcal{P}$ a class of semiparametric models.

Consider as if the data were generated from a conceptual class of parametric
models by substituting $\eta \left(  \cdot \right)  $ with a finite-dimensional
vector $\boldsymbol{\gamma}\in%
%TCIMACRO{\U{211d} }%
%BeginExpansion
\mathbb{R}
%EndExpansion
^{r}$ \cite{Newey1990}$:$\
\begin{equation}
\mathcal{P}_{\boldsymbol{\gamma}}^{sub}=\{p_{\mathbf{Z}}\left(  \mathbf{Z}%
_{\mathbf{i}};\boldsymbol{\beta},\boldsymbol{\gamma}\right)  ;\text{
}\boldsymbol{\beta}\in%
%TCIMACRO{\U{211d} }%
%BeginExpansion
\mathbb{R}
%EndExpansion
^{q},\text{ }\boldsymbol{\gamma}\in%
%TCIMACRO{\U{211d} }%
%BeginExpansion
\mathbb{R}
%EndExpansion
^{r}\} \subset \mathcal{P}\emph{,} \label{psub}%
\end{equation}
termed the parametric submodels. We restrict that at the truth, $p_{0}\left(
\mathbf{Z}_{\mathbf{i}};\boldsymbol{\theta}_{0}\right)  =p_{\mathbf{Z}}\left(
\mathbf{Z}_{\mathbf{i}};\boldsymbol{\beta}_{0},\boldsymbol{\gamma}_{0}\right)
=p_{\mathbf{Z}}\left(  \mathbf{Z}_{\mathbf{i}};\boldsymbol{\beta}_{0},\eta
_{0}(\cdot)\right)  $\ for some $\boldsymbol{\gamma}_{0}\in%
%TCIMACRO{\U{211d} }%
%BeginExpansion
\mathbb{R}
%EndExpansion
^{r}$. Let $\boldsymbol{\theta}^{sub}=$ $\left(  \boldsymbol{\beta}^{\top
},\boldsymbol{\gamma}^{\top}\right)  ^{\top}\in%
%TCIMACRO{\U{211d} }%
%BeginExpansion
\mathbb{R}
%EndExpansion
^{p}$ denote the parameter vector for the submodel where $p=q+r$.

Distinct from usual parametric models that characterize real study data with
model parameters, a parametric submodel is merely a \textquotedblleft
bridge\textquotedblright \ and not used to fit data since it requires
$p_{0}\left(  \mathbf{Z}_{\mathbf{i}};\boldsymbol{\theta}_{0}\right)
\in \mathcal{P}_{\boldsymbol{\gamma}}^{sub},$ where the truth is unknown.

In fact, an RAL estimator of $\boldsymbol{\beta}$ (the parameter of interest)
for a semiparametric model is also RAL for every parametric submodel
\cite{Tsaitis2006} in $\mathcal{P}_{\boldsymbol{\gamma}}^{sub}.$ But unlike
semiparametric models involving the infinite-dimensional $\eta(\cdot)$,
parametric submodels are granted the well-defined score vectors at the truth
$\boldsymbol{\theta}_{0}$:%
\begin{equation}
\mathbf{S}_{\boldsymbol{\theta}^{sub}}^{p\times1}\left(  \mathbf{Z}%
_{\mathbf{i}};\boldsymbol{\theta}_{0}\right)  =\left(  \mathbf{S}%
_{\boldsymbol{\beta}}^{\top}\left(  \mathbf{Z}_{\mathbf{i}};\boldsymbol{\theta
}_{0}\right)  ,\mathbf{S}_{\boldsymbol{\gamma}}^{\top}\left(  \mathbf{Z}%
_{\mathbf{i}};\boldsymbol{\theta}_{0}\right)  \right)  ^{\top},
\label{eqn.para.score}%
\end{equation}
where $\mathbf{S}_{\boldsymbol{\theta}^{sub}}\left(  \mathbf{Z}_{\mathbf{i}%
};\boldsymbol{\theta}_{0}\right)  =\partial \log p_{0}(\mathbf{Z}_{\mathbf{i}%
};\boldsymbol{\theta}_{0})/\partial \boldsymbol{\theta}^{sub\top}$, for
$\boldsymbol{\theta}^{sub}$ $($or $\boldsymbol{\beta},\boldsymbol{\gamma}).$

\subsection{Tangent Spaces\label{Ef.sec4.3 copy(1)}}

In $\mathcal{H}_{b}$ that consists of all measurable functions of
$\mathbf{h}^{q\times1}\left(  \mathbf{Z}_{\mathbf{i}}\right)  $ with mean zero
and finite variances, and equipped with both inner products b1\ and b2, the
score vectors for the submodels in (\ref{eqn.para.score}) can span linear
subspaces (with arbitrary matrix $\mathbf{B}$ of real numbers), termed
parametric submodel tangent spaces.

\subsubsection{Parametric Submodel Tangent Spaces}

\paragraph{Non-overlap Model Class 1}

\ The\ parametric submodel tangent space for\ model class 1 spanned by
$\mathbf{S}_{\boldsymbol{\theta}^{sub}}^{p\times1}\left(  \mathbf{Z}%
_{\mathbf{i}};\boldsymbol{\theta}_{0}\right)  $ is a linear subspace of
$\mathcal{H}_{b}$, where
\begin{align}
\pounds _{\boldsymbol{\beta \gamma}}^{sub}  &  =\{ \mathbf{BS}%
_{\boldsymbol{\theta}^{sub}}^{p\times1}\left(  \mathbf{Z}_{\mathbf{i}%
};\boldsymbol{\theta}_{0}\right)  ;\text{ }\forall \text{ arbitrary matrix
}\mathbf{B}^{q\times p}\}=\pounds _{\boldsymbol{\beta}}\oplus \Lambda
_{\boldsymbol{\gamma}},\text{ }\nonumber \\
\pounds _{\boldsymbol{\beta}}  &  =\left \{  \mathbf{BS}_{\boldsymbol{\beta}%
}^{q\times1}\left(  \mathbf{Z}_{\mathbf{i}};\boldsymbol{\theta}_{0}\right)
;\text{ }\forall \mathbf{B}^{q\times q}\right \}  ,\text{ }\Lambda
_{\boldsymbol{\gamma}}=\left \{  \mathbf{BS}_{\boldsymbol{\gamma}}^{r\times
1}\left(  \mathbf{Z}_{\mathbf{i}};\boldsymbol{\theta}_{0}\right)  ,\text{
}\forall \mathbf{B}^{q\times r}\right \}  , \label{eqn.ps.nts}%
\end{align}
with $\oplus$ denoting the direct sum. Since $\boldsymbol{\theta}^{sub}$ $=$
$\left(  \boldsymbol{\beta}^{\top},\boldsymbol{\gamma}^{\top}\right)  ^{\top
}\in%
%TCIMACRO{\U{211d} }%
%BeginExpansion
\mathbb{R}
%EndExpansion
^{p},$ $\pounds _{\boldsymbol{\beta \gamma}}^{sub}$\ is the direct sum of two
linear subspaces: $\pounds _{\boldsymbol{\beta}},$ the tangent space for
$\boldsymbol{\beta}$; and $\Lambda_{\boldsymbol{\gamma}}$, the tangent space
for $\boldsymbol{\gamma}$, also termed the parametric submodel nuisance
tangent space (submodel n.t.s.), $\Lambda_{\boldsymbol{\gamma}}$ is a key
component to find efficiency.

\paragraph{Enumerated Model Class 2}

\ For\ model class 2, the\ parametric submodel tangent space spanned by
$\mathcal{M}\left[  \mathbf{S}_{\boldsymbol{\theta}^{sub}}\left(
\mathbf{Z}_{\mathbf{i}};\boldsymbol{\theta}_{0}\right)  \right]  $ is
\begin{align*}
\widetilde{\pounds }_{\boldsymbol{\beta \gamma}}^{sub}  &  =\{ \mathbf{B}%
\mathcal{M}\left[  \mathbf{S}_{\boldsymbol{\theta}^{sub}}\left(
\mathbf{Z}_{\mathbf{i}};\boldsymbol{\theta}_{0}\right)  \right]  ;\text{
}\forall \text{ arbitrary matrix }\mathbf{B}^{q\times p}\}=\widetilde
{\pounds }_{\boldsymbol{\beta}}\oplus \widetilde{\Lambda}_{\boldsymbol{\gamma}%
},\\
\widetilde{\pounds }_{\boldsymbol{\beta}}  &  =\left \{  \mathbf{B}%
\mathcal{M}\left[  \mathbf{S}_{\boldsymbol{\beta}}^{q\times1}\left(
\mathbf{Z}_{\mathbf{i}};\boldsymbol{\theta}_{0}\right)  \right]  ;\text{
}\forall \mathbf{B}^{q\times q}\right \}  ,\text{ }\widetilde{\Lambda
}_{\boldsymbol{\gamma}}=\left \{  \mathbf{B}\mathcal{M}\left[  \mathbf{S}%
_{\boldsymbol{\gamma}}^{r\times1}\left(  \mathbf{Z}_{\mathbf{i}}%
;\boldsymbol{\theta}_{0}\right)  \right]  ,\text{ }\forall \mathbf{B}^{q\times
r}\right \}  ,
\end{align*}
where $\widetilde{\pounds }_{\boldsymbol{\beta \gamma}}^{sub},$ $\widetilde
{\pounds }_{\boldsymbol{\beta}}$ and $\widetilde{\Lambda}_{\boldsymbol{\gamma
}}$ are the respectively subspaces after mapping from $\mathcal{H}_{b}$ to
$\mathcal{H}_{w}$ using the U-statistics projection mapping\ $\mathcal{M}$ in
(\ref{U-mapping}). \ 

\subsubsection{Semiparametric Tangent Spaces \label{Ef.sec5.2}}

We are now in a position to extend the parametric submodel tangent spaces to
semiparametric tangent spaces, which is achieved with mean-square closure. As
$\boldsymbol{\beta}$ is unchanged for a semiparametric
model,\ $\pounds _{\boldsymbol{\beta}}$\ $\left(  \widetilde{\pounds }%
_{\boldsymbol{\beta}}\right)  $ remains the same, but the nuisance tangent
spaces need to accommodate the infinite-dimensional nuisance parameter for
semiparametric models. \  \ 

\paragraph{Non-overlap Model Class 1}

Let $\Upsilon$ be the collection of nuisance parameters $\boldsymbol{\gamma}$
for all possible parametric submodels in $\mathcal{P}_{\boldsymbol{\gamma}%
}^{sub}$\ defined in (\ref{psub}). Consider the unions of points
($\mathbf{h}\left(  \mathbf{Z}_{\mathbf{i}}\right)  $) in all the parametric
submodel nuisance tangent spaces $\Lambda_{\boldsymbol{\gamma}}$. With a
slight abuse of notation, we denote this union by $\Lambda^{\cup}%
=\cup_{\left \{  \boldsymbol{\gamma}\in \Upsilon \right \}  }\Lambda
_{\boldsymbol{\gamma}}.$

\textbf{Definition.} The semiparametric nuisance\ tangent\ space
(semiparametric n.t.s.) $\Lambda_{\eta}$ for\ model class 1 is the mean-square
closure (in terms of the norm b1) $\Lambda^{\cup}$.\ Namely, $\Lambda_{\eta}$
consists of all $\mathbf{h}\left(  \mathbf{Z}_{\mathbf{i}}\right)  \ $in
$\mathcal{H}_{b}$ for which there exists a sequence of $\mathbf{B}%
_{j}\mathbf{S}_{\boldsymbol{\gamma}_{j}}(\mathbf{Z}_{\mathbf{i}})$ $\in$
$\Lambda^{\cup}$ ($j=1,2,...$) such that \
\begin{equation}
\lim_{j\rightarrow \infty}\Vert \mathbf{h}^{q\times1}\left(  \mathbf{Z}%
_{\mathbf{i}}\right)  -\mathbf{B}_{j}^{q\times r_{j}}\mathbf{S}%
_{\boldsymbol{\gamma}_{j}}^{r_{j}\times1}(\mathbf{Z}_{\mathbf{i}})\Vert
_{b1}^{2}=0, \label{eqn.add01}%
\end{equation}
where $\mathbf{S}_{\boldsymbol{\gamma}_{j}}^{r_{j}\times1}(\mathbf{Z}%
_{\mathbf{i}})$ corresponds to a sequence of submodels characterized by
$\boldsymbol{\gamma}_{j}\in%
%TCIMACRO{\U{211d} }%
%BeginExpansion
\mathbb{R}
%EndExpansion
^{{\normalsize r_{j}}}.$ Each submodel and its associated dimension ($r_{j}$)
can vary with $j$, but $\Lambda_{\eta}$ covers all possibilities.\ 

Denote the entire semiparametric tangent space for class 1
by\ $\pounds =\pounds _{\beta}\oplus \Lambda_{\eta}$.\ 

\paragraph{Enumerated Model Class 2}

\textbf{Definition.} The semiparametric n.t.s. for\ model class 2, denoted by
$\widetilde{\Lambda}_{\eta}$, is the mean-square closure of $\widetilde
{\Lambda}^{\cup}=\cup_{\left \{  \boldsymbol{\gamma}\in \Upsilon \right \}
}\widetilde{\Lambda}_{\boldsymbol{\gamma}}$. It consists of all $\mathbf{h}%
\left(  \mathbf{Z}_{i}\right)  $ in $\mathcal{H}_{w}$\ which is either in
$\widetilde{\Lambda}^{\cup}$ or the limit of a convergent sequence
$\mathbf{h}_{j}\left(  \mathbf{Z}_{i}\right)  \in \widetilde{\Lambda}^{\cup}%
$\ ($j=1,2,...$), i.e., with the within-subject norm (\ref{eqn.w.01}), \
\begin{equation}
\lim_{j\rightarrow \infty}\Vert \mathbf{h}\left(  \mathbf{Z}_{i}\right)
-\mathbf{h}_{j}\left(  \mathbf{Z}_{i}\right)  \Vert_{w}^{2}=0. \label{gen.02}%
\end{equation}
The semiparametric tangent space for\ model class 2 is hence $\widetilde
{\pounds }=\widetilde{\pounds }_{\beta}\oplus \widetilde{\Lambda}_{\eta}.$

Both $\Lambda_{\eta}$ and $\widetilde{\Lambda}_{\eta}$ are closed spaces by
definition. The theorem below shows that linearity and closedness are
preserved under the projection mapping $\mathcal{M}$.

\textbf{Theorem 4.} \ The semiparametric n.t.s. $\Lambda_{\eta}$ and
$\widetilde{\Lambda}_{\eta}$ are both linear subspaces and $\widetilde
{\Lambda}_{\eta}=\mathcal{M}\left(  {\normalsize \Lambda_{\eta}}\right)  $.

Therefore, all $h\left(  \mathbf{Z}_{\mathbf{i}}\right)  \in \mathcal{H}_{b}$
has a unique projection (up to its equivalence class)\ onto semiparametric
n.t.s. $\Lambda_{\eta}$ and $\widetilde{\Lambda}_{\eta}$. \ 

\subsection{Dual Geometric Interpretations for Semiparametric
Models\label{Ef.sec6.4}}

We now introduce a fundamental connection for the two classes of models termed
dual orthogonality, which is a direct generalization of properties for
parametric models by leveraging the submodel bridge. It geometrically
characterizes the semiparametric RAL estimators through influence functions
and semiparametric nuisance tangent spaces.

\textbf{Theorem 5.} \ A semiparametric RAL estimator of $\boldsymbol{\beta}$
for either class of models must have an influence function (I.F.)
$\boldsymbol{\varphi}\left(  \mathbf{Z}_{\mathbf{i}}\right)  $ satisfying
\begin{align}
\text{(i)}  &  :\left \langle \boldsymbol{\varphi}\left(  \mathbf{Z}%
_{\mathbf{i}}\right)  ,\mathbf{S}_{\boldsymbol{\beta}}\left(  \mathbf{Z}%
_{\mathbf{i}};\boldsymbol{\theta}_{0}\right)  \right \rangle _{b1}=E\left[
\boldsymbol{\varphi}\left(  \mathbf{Z}_{\mathbf{i}}\right)  \mathbf{S}%
_{\boldsymbol{\beta}}^{\top}\left(  \mathbf{Z}_{\mathbf{i}};\boldsymbol{\theta
}_{0}\right)  \right]  =\mathbf{I}_{q}\mathbf{,}\text{ }\label{eqn.semi}\\
\text{(ii)}  &  :\Pi_{b1}\left \{  \boldsymbol{\varphi}\left(  \mathbf{Z}%
_{\mathbf{i}}\right)  \mid \Lambda_{\eta}\right \}  =\mathbf{0},\nonumber \\
\text{(iii)}  &  :\Pi_{b2}\left \{  \boldsymbol{\varphi}\left(  \mathbf{Z}%
_{\mathbf{i}}\right)  \mid \widetilde{\Lambda}_{\eta}\right \}  =\Pi_{w}\left \{
2E\left[  \boldsymbol{\varphi}\left(  \mathbf{Z}_{\mathbf{i}}\right)
\mid \mathbf{Z}_{i_{1}}\right]  \mid \widetilde{\Lambda}_{\eta}\right \}
=\mathbf{0},\nonumber
\end{align}
where $\mathbf{I}_{q}$ is the $q\times q$ identity matrix, $\Pi_{b1}\left \{
\boldsymbol{\varphi}\left(  \mathbf{Z}_{\mathbf{i}}\right)  \mid \Lambda_{\eta
}\right \}  $\ denotes the unique projection (w.r.t. inner product 1) of
$\boldsymbol{\varphi}\left(  \mathbf{Z}_{\mathbf{i}}\right)  $ onto
$\Lambda_{\eta}$, $\Pi_{b2}\left \{  \boldsymbol{\varphi}\left(  \mathbf{Z}%
_{\mathbf{i}}\right)  \mid \widetilde{\Lambda}_{\eta}\right \}  $\ is the unique
projection (w.r.t. inner product 2) of $\boldsymbol{\varphi}\left(
\mathbf{Z}_{\mathbf{i}}\right)  $ onto $\widetilde{\Lambda}_{\eta}$. (ii) and
(iii) imply that $\boldsymbol{\varphi}\left(  \mathbf{Z}_{\mathbf{i}}\right)
$ is deemed dual orthogonal to both the semiparametric n.t.s. $\Lambda_{\eta}$
(for model class 1) and its mapping $\widetilde{\Lambda}_{\eta}$ (for model
class 2)$.$

While Theorem 3 asserts that the two classes of models share the same
RAL\ estimators and I.F.s., Theorem 5 further identifies such estimators
through the dual orthogonality between I.F.s and respective semiparametric
n.t.s. $\Lambda_{\eta}$ and $\widetilde{\Lambda}_{\eta}$. Recall that the
variance of any element is always larger than or equal to its projection onto
a linear subspace in (\ref{Var1}) and (\ref{Var2}). This intrinsic connection
between the two model classes allows us to locate the efficient estimator for
model class 2 through model class 1, which serves a \textquotedblleft
conjugate\textquotedblright \ model class.

\section{Semiparametric Efficiency Bound\label{New.Sec6}}

Again, we aim to to identify the efficient semiparametric RAL estimator for
the FRM in (\ref{eqn 2.15}), or the model class 2.\ Directly tackling the
efficiency for class 2 is difficult, but the dual orthogonality motivates a
strategy to find efficient estimator via model class 1, which is more
straightforward. Due to the many-to-one mapping $\mathcal{M}$, the efficient
I.F. for class 2 corresponds to multiple I.F.s in class 1, but our goal is
fulfilled if we identify one of them in the equivalence class. We start by
establishing the efficient I.F. for model class 1.

\textbf{Definition.} The efficient I.F. is the unique influence function (up
to its equivalence class) belonging to the tangent space that has the smallest
asymptotic variance.

Recall that a semiparametric RAL estimator of\ $\boldsymbol{\beta}$ in
$\mathcal{P}$ is an RAL estimator for every parametric submodel. In terms of
influence functions, the class of I.F.s for a semiparametric model will be a
subset of the class of I.F.s for all parametric submodels. Any semiparametric
influence function must be orthogonal to all parametric submodel nuisance
tangent spaces. Hence, the asymptotic variance of a semiparametric model must
be greater than or equal to the parametric efficiency bound for any submodel,
or the supremum of such bounds for all submodels.

We define the efficiency bound via this bridge for each model class.

\subsection{Parametric Submodels}

\subsubsection{Non-overlap Model Class 1}

The efficient I.F. for a parametric submodel in\ class 1, denoted by
$\boldsymbol{\varphi}_{\boldsymbol{\gamma}\text{,eff1}}^{sub}\left(
\mathbf{Z}_{\mathbf{i}};\boldsymbol{\theta}_{0}\right)  $,\ is the unique I.F.
in the tangent space $\pounds _{\beta \boldsymbol{\gamma}}^{sub}%
=\pounds _{\beta}\oplus \Lambda_{\boldsymbol{\gamma}}$ with the smallest norm
b1, i.e., for any I.F. $\boldsymbol{\varphi}^{sub}\left(  \mathbf{Z}%
_{\mathbf{i}};\boldsymbol{\theta}_{0}\right)  $ of a submodel in
$\mathcal{P}_{\boldsymbol{\gamma}}^{sub}$,
\[
\left \Vert \boldsymbol{\varphi}_{\boldsymbol{\gamma}\text{,eff1}}^{sub}\left(
\mathbf{Z}_{\mathbf{i}};\boldsymbol{\theta}_{0}\right)  \right \Vert _{b1}%
^{2}\leq \left \Vert \boldsymbol{\varphi}^{sub}\left(  \mathbf{Z}_{\mathbf{i}%
};\boldsymbol{\theta}_{0}\right)  \right \Vert _{b1}^{2},\text{ }%
\boldsymbol{\varphi}_{\boldsymbol{\gamma}\text{,eff1}}^{sub}\left(
\mathbf{Z}_{\mathbf{i}};\boldsymbol{\theta}_{0}\right)  =\Pi_{b1}\left \{
\boldsymbol{\varphi}^{sub}\left(  \mathbf{Z}_{\mathbf{i}};\boldsymbol{\theta
}_{0}\right)  \mid \pounds _{\beta \boldsymbol{\gamma}}^{sub}\right \}  ,
\]
then the efficiency bound for parametric submodels of class 1 is its variance
\[
\upsilon_{1,\boldsymbol{\gamma}}^{sub}=Var\left[  \boldsymbol{\varphi
}_{\boldsymbol{\gamma}\text{,eff1}}^{sub}\left(  \mathbf{Z}_{\mathbf{i}%
};\boldsymbol{\theta}_{0}\right)  \right]  .
\]
The semiparametric efficiency bound\ for model class 1 is the supremum of
$\upsilon_{1,\boldsymbol{\gamma}}^{sub}$ over all submodels:
\begin{equation}
\upsilon_{1}=\sup_{\left \{  \mathcal{P}_{\boldsymbol{\gamma}}^{sub}\right \}
}\upsilon_{1,\boldsymbol{\gamma}}^{sub}=\sup_{\left \{  \mathcal{P}%
_{\boldsymbol{\gamma}}^{sub}\right \}  }Var\left[  \boldsymbol{\varphi
}_{\boldsymbol{\gamma}\text{,eff1}}^{sub}\left(  \mathbf{Z}_{\mathbf{i}%
};\boldsymbol{\theta}_{0}\right)  \right]  , \label{def.11}%
\end{equation}
where $\sup$ is defined based on the non-negative definite criterion
to\ compare matrices using their differences. \ 

\subsubsection{Enumerated Model Class 2\ }

Likewise, the\ efficient I.F. for parametric submodels in\ class 2,
$\boldsymbol{\psi}_{\boldsymbol{\gamma}\text{,eff2}}^{sub}\left(
\mathbf{Z}_{\mathbf{i}};\boldsymbol{\theta}_{0}\right)  $, is the I.F. lying
in $\pounds _{\beta \boldsymbol{\gamma}}^{sub}=\pounds _{\beta}\oplus
\Lambda_{\boldsymbol{\gamma}}$ with the smallest norm b2. Hence, any I.F.
$\boldsymbol{\psi}^{sub}\left(  \mathbf{Z}_{\mathbf{i}};\boldsymbol{\theta
}_{0}\right)  $ of a submodel satisfies%
\[
\left \Vert \boldsymbol{\psi}_{\boldsymbol{\gamma}\text{,eff2}}^{sub}\left(
\mathbf{Z}_{\mathbf{i}};\boldsymbol{\theta}_{0}\right)  \right \Vert _{b2}%
^{2}=\left \Vert \mathcal{M}\left[  \boldsymbol{\psi}_{\boldsymbol{\gamma
}\text{,eff2}}^{sub}\left(  \mathbf{Z}_{\mathbf{i}};\boldsymbol{\theta}%
_{0}\right)  \right]  \right \Vert _{w}^{2}\leq \left \Vert \mathcal{M}\left[
\boldsymbol{\psi}^{sub}\left(  \mathbf{Z}_{\mathbf{i}};\boldsymbol{\theta}%
_{0}\right)  \right]  \right \Vert _{w}^{2}=\left \Vert \boldsymbol{\psi}%
^{sub}\left(  \mathbf{Z}_{\mathbf{i}};\boldsymbol{\theta}_{0}\right)
\right \Vert _{b2}^{2}.
\]

By the multivariate Pythagoras, any two I.F.s with zero difference in norm b2
are equivalent as they determine the same efficiency (asymptotic variance).
The equivalence class for $\boldsymbol{\psi}_{\boldsymbol{\gamma}\text{,eff2}%
}^{sub}\left(  \mathbf{Z}_{\mathbf{i}};\boldsymbol{\theta}_{0}\right)  $ is
hence defined to be
\begin{equation}
\Gamma_{\text{eff2}}^{sub}=\left \{  \boldsymbol{\psi}^{sub}\left(
\mathbf{Z}_{\mathbf{i}};\boldsymbol{\theta}_{0}\right)  \in \pounds _{\beta
\boldsymbol{\gamma}}^{sub}:\mathcal{M}\left[  \boldsymbol{\psi}^{sub}\left(
\mathbf{Z}_{\mathbf{i}};\boldsymbol{\theta}_{0}\right)  \right]
=\mathcal{M}\left[  \boldsymbol{\psi}_{\boldsymbol{\gamma}\text{,eff2}}%
^{sub}\left(  \mathbf{Z}_{\mathbf{i}};\boldsymbol{\theta}_{0}\right)  \right]
\text{ a.s.}\right \}  , \label{eqn.equiv.01}%
\end{equation}
$\boldsymbol{\psi}_{\boldsymbol{\gamma}\text{,eff2}}^{sub}\left(
\mathbf{Z}_{\mathbf{i}};\boldsymbol{\theta}_{0}\right)  $ is unique up to this
equivalence class $\Gamma_{\text{eff2}}^{sub}$. The efficiency bound for
parametric\ submodels $\upsilon_{2,\boldsymbol{\gamma}}^{sub}$ and the
semiparametric efficiency bound for class 2 $\upsilon_{2}$ are respectively
defined by
\[
\upsilon_{2,\boldsymbol{\gamma}}^{sub}=Var\left \{  \mathcal{M}\left[
\boldsymbol{\psi}_{\boldsymbol{\gamma}\text{,eff2}}^{sub}\left(
\mathbf{Z}_{\mathbf{i}};\boldsymbol{\theta}_{0}\right)  \right]  \right \}
\text{, }\upsilon_{2}=\sup_{\left \{  \mathcal{P}_{\gamma}^{sub}\right \}
}\upsilon_{2,\boldsymbol{\gamma}}^{sub}.
\]
The theorem below connects the two classes of submodels regarding the
efficient I.F..

\textbf{Theorem 6.} \ The norm b2 of the efficient I.F. for class 1,
$\boldsymbol{\varphi}_{\boldsymbol{\gamma}\text{,eff1}}^{sub}\left(
\mathbf{Z}_{\mathbf{i}};\boldsymbol{\theta}_{0}\right)  ,$ equals the norm b2
of the efficient I.F. for class 2, $\boldsymbol{\psi}_{\boldsymbol{\gamma
}\text{,eff2}}^{sub}\left(  \mathbf{Z}_{\mathbf{i}};\boldsymbol{\theta}%
_{0}\right)  ,$ hence is in the equivalence class $\Gamma_{\text{eff2}}^{sub}$
defined in (\ref{eqn.equiv.01}), i.e.,
\[
\left \Vert \boldsymbol{\varphi}_{\boldsymbol{\gamma}\text{,eff1}}^{sub}\left(
\mathbf{Z}_{\mathbf{i}};\boldsymbol{\theta}_{0}\right)  \right \Vert _{b2}%
^{2}=\left \Vert \boldsymbol{\psi}_{\boldsymbol{\gamma}\text{,eff2}}%
^{sub}\left(  \mathbf{Z}_{\mathbf{i}};\boldsymbol{\theta}_{0}\right)
\right \Vert _{b2}^{2}\text{, or }\left \Vert \mathcal{M}\left[
\boldsymbol{\varphi}_{\boldsymbol{\gamma}\text{,eff1}}^{sub}\left(
\mathbf{Z}_{\mathbf{i}};\boldsymbol{\theta}_{0}\right)  \right]  \right \Vert
_{w}^{2}=\left \Vert \mathcal{M}\left[  \boldsymbol{\psi}_{\boldsymbol{\gamma
}\text{,eff2}}^{sub}\left(  \mathbf{Z}_{\mathbf{i}};\boldsymbol{\theta}%
_{0}\right)  \right]  \right \Vert _{w}^{2}.
\]

By Theorem 3, $\boldsymbol{\varphi}_{\boldsymbol{\gamma}\text{,eff1}}%
^{sub}\left(  \mathbf{Z}_{\mathbf{i}};\boldsymbol{\theta}_{0}\right)  $ is
already a valid I.F. for model 2, now admitting the same norm b2 as
$\boldsymbol{\psi}_{\boldsymbol{\gamma}\text{,eff2}}^{sub}\left(
\mathbf{Z}_{\mathbf{i}};\boldsymbol{\theta}_{0}\right)  $, it is indeed in the
equivalence class $\Gamma_{\text{eff2}}^{sub}$. It follows from Theorem 6 and
the definition of norm b2 that after mapping, $\mathcal{M}\left[
\boldsymbol{\varphi}_{\boldsymbol{\gamma}\text{,eff1}}^{sub}\left(
\mathbf{Z}_{\mathbf{i}};\boldsymbol{\theta}_{0}\right)  \right]
=\mathcal{M}\left[  \boldsymbol{\psi}_{\boldsymbol{\gamma}\text{,eff2}}%
^{sub}\left(  \mathbf{Z}_{\mathbf{i}};\boldsymbol{\theta}_{0}\right)  \right]
$ a.s., hence they determine the same asymptotic variance.

Not surprisingly, $\boldsymbol{\varphi}_{\boldsymbol{\gamma}\text{,eff1}%
}^{sub}\left(  \mathbf{Z}_{\mathbf{i}};\boldsymbol{\theta}_{0}\right)  $
delivers exactly what we aim to find: one element lying in the submodel
tangent space $\pounds _{\beta \boldsymbol{\gamma}}^{sub}$ that yields the
smallest variance for model class 2.

\subsection{Semiparametric Models}

Now we switch from parametric submodels to semiparametric models by
considering the semiparametric n.t.s. $\Lambda_{\eta}$ defined in
(\ref{eqn.add01}). To differentiate from those of submodels in notation, we
drop the superscripts (of \textquotedblleft sub\textquotedblright) and
subscripts $\boldsymbol{\gamma}$ for quantities of semiparametric models. By
definition, the\ efficient I.F. for the\ semiparametric model 1 is the I.F. in
$\pounds $ whose variance achieves the semiparametric efficiency bound
$\upsilon_{1}$. With variationally independent parameters $\boldsymbol{\theta
}=\{ \boldsymbol{\beta},\eta(\cdot)\}$, an impoartant result is that the
efficient score is the residual of the score vector for $\boldsymbol{\beta}$
after projecting it onto the nuisance tangent space \cite{Tsaitis2006}$.$ For
model class 1, the\ semiparametric efficient score is easily found to be
\begin{equation}
\mathbf{S}_{\text{eff1}}\left(  \mathbf{Z}_{\mathbf{i}};\boldsymbol{\theta
}_{0}\right)  =\mathbf{S}_{\boldsymbol{\beta}}\left(  \mathbf{Z}_{\mathbf{i}%
};\boldsymbol{\theta}_{0}\right)  -\Pi_{b1}\left \{  \mathbf{S}%
_{\boldsymbol{\beta}}\left(  \mathbf{Z}_{\mathbf{i}};\boldsymbol{\theta}%
_{0}\right)  \mid \Lambda_{\eta}\right \}  . \label{eqn.eff.score}%
\end{equation}
Below shows how to find the efficient I.F. from this score.

\textbf{Theorem 7.}\ Let
\begin{equation}
\boldsymbol{\varphi}_{\text{eff1}}\left(  \mathbf{Z}_{\mathbf{i}%
};\boldsymbol{\theta}_{0}\right)  =E^{-1}\left(  \mathbf{S}_{\text{eff1}%
}\mathbf{S}_{\text{eff1}}^{\top}\right)  \mathbf{S}_{\text{eff1}}\left(
\mathbf{Z}_{\mathbf{i}};\boldsymbol{\theta}_{0}\right)  . \label{Thm9}%
\end{equation}
Then $\boldsymbol{\varphi}_{\text{eff1}}\left(  \mathbf{Z}_{\mathbf{i}%
};\boldsymbol{\theta}_{0}\right)  $ is the unique element in
$\pounds =\pounds _{\beta}\oplus \Lambda_{\eta}$ whose variance achieves
$\upsilon_{1}.$ \ 

Akin to submodels, this semiparametric efficient I.F. $\boldsymbol{\varphi
}_{\text{eff1}}\left(  \mathbf{Z}_{\mathbf{i}};\boldsymbol{\theta}_{0}\right)
$ for model class 1 is also mapped to an element in $\widetilde
{\mathbf{\pounds }}$ that achieves the efficiency for the model class 2,
summarized in the following theorem that generalizes Theorem 6 for submodels.

\textbf{Theorem 8.} Let $\boldsymbol{\psi}_{\text{eff2}}\left(  \mathbf{Z}%
_{\mathbf{i}};\boldsymbol{\theta}_{0}\right)  $ denote the efficient I.F. for
the semiparametric model class 2. $\boldsymbol{\varphi}_{\text{eff1}}\left(
\mathbf{Z}_{\mathbf{i}};\boldsymbol{\theta}_{0}\right)  $ has the same norm b2
as $\boldsymbol{\psi}_{\text{eff2}}\left(  \mathbf{Z}_{\mathbf{i}%
};\boldsymbol{\theta}_{0}\right)  $ and hence is in its equivalence class
denoted by $\Gamma_{\text{eff2}}$, i.e.,%
\[
\left \Vert \boldsymbol{\varphi}_{\text{eff1}}\left(  \mathbf{Z}_{\mathbf{i}%
};\boldsymbol{\theta}_{0}\right)  \right \Vert _{b2}^{2}=\left \Vert
\boldsymbol{\psi}_{\text{eff2}}\left(  \mathbf{Z}_{\mathbf{i}}%
;\boldsymbol{\theta}_{0}\right)  \right \Vert _{b2}^{2}\text{, or }\left \Vert
\mathcal{M}\left[  \boldsymbol{\varphi}_{\text{eff1}}\left(  \mathbf{Z}%
_{\mathbf{i}};\boldsymbol{\theta}_{0}\right)  \right]  \right \Vert _{w}%
^{2}=\left \Vert \mathcal{M}\left[  \boldsymbol{\psi}_{\text{eff2}}\left(
\mathbf{Z}_{\mathbf{i}};\boldsymbol{\theta}_{0}\right)  \right]  \right \Vert
_{w}^{2},
\]
where $\Gamma_{\text{eff2}}=\left \{  \boldsymbol{\psi}\left(  \mathbf{Z}%
_{\mathbf{i}};\boldsymbol{\theta}_{0}\right)  \in \pounds :\mathcal{M}\left[
\boldsymbol{\psi}\left(  \mathbf{Z}_{\mathbf{i}};\boldsymbol{\theta}%
_{0}\right)  \right]  =\mathcal{M}\left[  \boldsymbol{\psi}_{\text{eff2}%
}\left(  \mathbf{Z}_{\mathbf{i}};\boldsymbol{\theta}_{0}\right)  \right]
\text{ a.s.}\right \}  $. \ 

Again, the multivariate Pythagoras implies that the variance of $\mathcal{M}%
\left[  \boldsymbol{\psi}_{\text{eff2}}\left(  \mathbf{Z}_{\mathbf{i}%
};\boldsymbol{\theta}_{0}\right)  \right]  $equals the semiparametric
efficiency bound $\upsilon_{2}$\ for model class 2, as a direct result of
Theorem 8, we have\
\[
Var\left \{  \mathcal{M}\left[  \boldsymbol{\varphi}_{\text{eff1}}\left(
\mathbf{Z}_{\mathbf{i}};\boldsymbol{\theta}_{0}\right)  \right]  \right \}
=Var\left \{  \mathcal{M}\left[  \boldsymbol{\psi}_{\text{eff2}}\left(
\mathbf{Z}_{\mathbf{i}};\boldsymbol{\theta}_{0}\right)  \right]  \right \}
=\upsilon_{2}.
\]

Based on Theorem 8, we can identify the efficient estimator for model class 2
via model class 1, where constructing the efficient I.F. $\boldsymbol{\varphi
}_{\text{eff1}}\left(  \mathbf{Z}_{\mathbf{i}};\boldsymbol{\theta}_{0}\right)
$ is apparently more straightforward. We now apply those results to the
regression setting of FRM.

\section{The Efficiency for the FRM\label{New.sec5}}

By the conditions in Theorem 5 that any I.F. satisfies, to derive the
semiparametric efficient I.F. $\boldsymbol{\varphi}_{\text{eff1}}\left(
\mathbf{Z}_{\mathbf{i}};\boldsymbol{\theta}_{0}\right)  $ for model class 1,
we first identify the form of the semiparametric n.t.s. $\Lambda_{\eta}$ and
then find elements that are orthogonal to it. These elements in the orthogonal
complement $\Lambda_{\eta}^{\bot}$ form a pool of candidates for the optimal one.

Consider $\mathbf{Z}_{\mathbf{i}}=(\mathbf{Y}_{\mathbf{i}}^{\top}%
,\mathbf{X}_{\mathbf{i}}^{\top})^{\top}$, $\mathbf{Y}_{\mathbf{i}}%
=(\mathbf{Y}_{i_{1}}^{\top}\mathbf{,Y}_{i_{2}}^{\top})^{\top}$, $\mathbf{X}%
_{\mathbf{i}}\mathbf{=(X}_{i_{1}}^{\top}\mathbf{,X}_{i_{2}}^{\top}%
\mathbf{)}^{\top},$ $\mathbf{i}=\left(  i_{1},i_{2}\right)  \in C_{2}^{n},$
where $X_{i}$\textbf{ }$(Y_{i})$ is a $q\times1$ $(m\times1)$ vector of
explanatory variables (outcomes) for the $i$-th subject. Let $f_{\mathbf{i}%
}(\mathbf{Y}_{i_{1}}\mathbf{,Y}_{i_{2}})$ be a univariate continuous response
for the $i$-th pair such as the microbiome Beta-diversity in (\ref{eqn.n1})
(same considerations apply to more general response types \cite{Tsaitis2006}%
)$.$ The semiparametric FRM for a continuous $f_{\mathbf{i}}$ is\
\begin{equation}
f_{\mathbf{i}}=h\left(  \mathbf{X}_{\mathbf{i}};\boldsymbol{\beta}\right)
+\varepsilon_{\mathbf{i}},\text{ }E(\varepsilon_{\mathbf{i}}\mid
\mathbf{X}_{\mathbf{i}})=0, \label{EQN.frm.01}%
\end{equation}
where $\varepsilon_{\mathbf{i}}=f_{\mathbf{i}}-h\left(  \mathbf{X}%
_{\mathbf{i}},\boldsymbol{\beta}\right)  $ is the residual of between-subject attributes.

The goal is to identify the semiparametric RAL estimator of $\boldsymbol{\beta
}$ with the smallest variance for the FRM, but we can tackle it by finding
$\boldsymbol{\varphi}_{\text{eff1}}\left(  \varepsilon_{\mathbf{i}}%
,\mathbf{X}_{\mathbf{i}};\boldsymbol{\theta}_{0}\right)  $ for model class 1
first$.$

\subsection{Identifying $\Lambda_{\eta}$ with the Joint Likelihood and
Score\label{Ef.sec6.2}}

The joint density of observed between-subject attributes $(\varepsilon
_{\mathbf{i}},\mathbf{X}_{\mathbf{i}}),$ belongs to a class of semiparametric
models
\begin{equation}
\mathcal{P}=\left \{  p_{\varepsilon,\mathbf{X}}\left(  \varepsilon
_{\mathbf{i}},\mathbf{X}_{\mathbf{i}};\boldsymbol{\beta},\eta \left(
\cdot \right)  \right)  ;\text{ }\boldsymbol{\beta}\in%
%TCIMACRO{\U{211d} }%
%BeginExpansion
\mathbb{R}
%EndExpansion
^{q}\text{ and }\eta \left(  \cdot \right)  \  \text{is infinite-dimensional}%
\right \}  . \label{eqn.n7n}%
\end{equation}
We assume that the underlying true data are generated from $p\left(
\mathbf{Y}_{i},\mathbf{X}_{i};\boldsymbol{\theta}_{0}\right)  $, which induces
$p\left(  \mathbf{Y}_{\mathbf{i}},\mathbf{X}_{\mathbf{i}};\boldsymbol{\theta
}_{0}\right)  =p\left(  \mathbf{Y}_{i_{1}},\mathbf{X}_{i_{1}}\right)  p\left(
\mathbf{Y}_{i_{2}},\mathbf{X}_{i_{2}}\right)  .$ By independence and the
change of variables, $\boldsymbol{\theta}_{0}$ remains the same for describing
the individual-level $p\left(  \mathbf{Y}_{i},\mathbf{X}_{i}\right)  $ and
pairwise-level $p\left(  \mathbf{Y}_{\mathbf{i}},\mathbf{X}_{\mathbf{i}%
}\right)  $ or $p\left(  \varepsilon_{\mathbf{i}},\mathbf{X}_{\mathbf{i}%
}\right)  $. We denote the truth by $p_{\varepsilon,\mathbf{X}}\left(
\varepsilon_{\mathbf{i}},\mathbf{X}_{\mathbf{i}};\boldsymbol{\theta}%
_{0}\right)  .$ The\ parametric submodels for $P$are given by
\[
\mathcal{P}_{\boldsymbol{\gamma}}^{sub}=\{p_{\varepsilon,\mathbf{X}}\left(
\varepsilon_{\mathbf{i}},\mathbf{X}_{\mathbf{i}};\boldsymbol{\beta
},\boldsymbol{\gamma}\right)  ;\text{ }\boldsymbol{\beta}\in%
%TCIMACRO{\U{211d} }%
%BeginExpansion
\mathbb{R}
%EndExpansion
^{q},\text{ }\boldsymbol{\gamma}\in%
%TCIMACRO{\U{211d} }%
%BeginExpansion
\mathbb{R}
%EndExpansion
^{r}\} \subset \mathcal{P},
\]
which contain the truth\ $\boldsymbol{\theta}_{0}=\{ \boldsymbol{\beta}%
_{0},\eta_{0}(\cdot)\}$. Let $\Lambda_{\eta}$ denote the semiparametric n.t.s.
for model class 1 resulting from the mean-square closure of $\Lambda^{\cup
}=\cup_{\left \{  \boldsymbol{\gamma}\in \Upsilon \right \}  }\Lambda
_{\boldsymbol{\gamma}},$ the union of all parametric submodels n.t.s.. We can
readily determine the form of $\Lambda_{\eta}$ by applying arguments similar
to those for the classical within-subject semiparametric models
\cite{Tsaitis2006}$,$ summarized below.

\textbf{Theorem 9. }\ The space $\Lambda_{\eta}$ contains all mean-zero
functions $\boldsymbol{\lambda}\left(  \varepsilon_{\mathbf{i}},\mathbf{X}%
_{\mathbf{i}}\right)  $ satisfying the constraint on the conditional mean in
(\ref{EQN.frm.01}), namely,
\begin{equation}
\Lambda_{\eta}=\left \{  \boldsymbol{\lambda}^{q\times1}\left(  \varepsilon
_{\mathbf{i}},\mathbf{X}_{\mathbf{i}}\right)  :E\left[  \boldsymbol{\lambda
}\left(  \varepsilon_{\mathbf{i}},\mathbf{X}_{\mathbf{i}}\right)
\varepsilon_{\mathbf{i}}\mid \mathbf{X}_{\mathbf{i}}\right]  =\mathbf{0}%
^{q\times1}\right \}  . \label{eqn.n.t.s.}%
\end{equation}
Its orthogonal complement (w.r.t. inner product b1) is defined by%
\[
\Lambda_{\eta}^{\bot}=\left \{  \boldsymbol{\chi}^{q\times1}\left(
\varepsilon_{\mathbf{i}},\mathbf{X}_{\mathbf{i}}\right)  \in \mathcal{H}%
_{b}:\left \langle \boldsymbol{\chi}\left(  \varepsilon_{\mathbf{i}}%
,\mathbf{X}_{\mathbf{i}}\right)  ,\boldsymbol{\lambda}\left(  \varepsilon
_{\mathbf{i}},\mathbf{X}_{\mathbf{i}}\right)  \right \rangle _{b1}=0\right \}
.
\]

The form of $\Lambda_{\eta}$ for model class 1 above is conformable with that
for the semiparametric GLM \cite{Tsaitis2006}$,$ as both models have
restrictions only on the conditional mean.

\subsection{The Efficient Influence Function of the FRM \label{Ef.sec6.6}}

Recall that the efficient score is the residual after projecting
$\mathbf{S}_{\boldsymbol{\beta}}$ onto $\Lambda_{\eta}$ by
(\ref{eqn.eff.score}). In $\mathcal{H}_{b}$, the projection (w.r.t. inner
product 1) of an arbitrary element $\boldsymbol{g}\left(  \varepsilon
_{\mathbf{i}},\mathbf{X}_{\mathbf{i}}\right)  \in \mathcal{H}_{b}$ onto
$\Lambda_{\eta}$ is readily shown to satisfy:%
\begin{equation}
\Pi_{b1}\left \{  \boldsymbol{g}\left(  \varepsilon_{\mathbf{i}},\mathbf{X}%
_{\mathbf{i}}\right)  \mid \Lambda_{\eta}^{\bot}\right \}  =\boldsymbol{g}%
-\Pi_{b1}\left \{  \boldsymbol{g}\left(  \varepsilon_{\mathbf{i}}%
,\mathbf{X}_{\mathbf{i}}\right)  \mid \Lambda_{\eta}\right \}  =E\left[
\boldsymbol{g}\left(  \varepsilon_{\mathbf{i}},\mathbf{X}_{\mathbf{i}}\right)
\varepsilon_{\mathbf{i}}\mid \mathbf{X}_{\mathbf{i}}\right]  E^{-1}%
(\varepsilon_{\mathbf{i}}^{2}\mid \mathbf{X}_{\mathbf{i}})\varepsilon
_{\mathbf{i}}, \label{proj.01}%
\end{equation}
which is verified by the fact that $\left \langle \Pi_{b1}\left \{
\boldsymbol{g}\mid \Lambda_{\eta}^{\bot}\right \}  ,\boldsymbol{\lambda}^{\ast
}\left(  \varepsilon_{\mathbf{i}},\mathbf{X}_{\mathbf{i}}\right)
\right \rangle _{b1}=0$ for any $\boldsymbol{\lambda}^{\ast}\left(
\varepsilon,\mathbf{X}\right)  \in \Lambda_{\eta}.$ Substituting $\mathbf{S}%
_{\boldsymbol{\beta}}\left(  \varepsilon_{\mathbf{i}},\mathbf{X}_{\mathbf{i}%
}\right)  $ in place of $\boldsymbol{g}\left(  \varepsilon_{\mathbf{i}%
},\mathbf{X}_{\mathbf{i}}\right)  $ in (\ref{proj.01}) yields the efficient
score for model 1:
\begin{equation}
\mathbf{S}_{\text{eff1}}\left(  \varepsilon_{\mathbf{i}},\mathbf{X}%
_{\mathbf{i}};\boldsymbol{\theta}_{0}\right)  =\mathbf{S}_{\boldsymbol{\beta}%
}-\Pi_{b1}\left \{  \mathbf{S}_{\boldsymbol{\beta}}\left(  \varepsilon
_{\mathbf{i}},\mathbf{X}_{\mathbf{i}}\right)  \mid \Lambda_{\eta}\right \}
=E\left[  \mathbf{S}_{\boldsymbol{\beta}}\left(  \varepsilon_{\mathbf{i}%
},\mathbf{X}_{\mathbf{i}}\right)  \varepsilon_{\mathbf{i}}\mid \mathbf{X}%
_{\mathbf{i}}\right]  V^{-1}\left(  \mathbf{X}_{\mathbf{i}}\right)
\varepsilon_{\mathbf{i}}, \label{eqn.n18}%
\end{equation}
where $V\left(  \mathbf{X}_{\mathbf{i}}\right)  =E\left(  \varepsilon
_{\mathbf{i}}^{2}\mid \mathbf{X}_{\mathbf{i}}\right)  $. By fixing $\eta
(\cdot)$ at the truth $\eta_{0}(\cdot)$ and taking partial derivatives w.r.t.
$\boldsymbol{\beta}$ of the conditional mean restriction $E\left[
f_{\mathbf{i}}-h\left(  \mathbf{X}_{\mathbf{i}};\boldsymbol{\beta}\right)
\mid \mathbf{X}_{\mathbf{i}}\right]  =0,$ we obtain \
\begin{equation}
E\left[  \varepsilon_{\mathbf{i}}\mathbf{S}_{\boldsymbol{\beta}}^{\top}\left(
\varepsilon_{\mathbf{i}},\mathbf{X}_{\mathbf{i}}\right)  \mid \mathbf{X}%
_{\mathbf{i}}\right]  =\frac{\partial}{\partial \boldsymbol{\beta}^{\top}%
}h\left(  \mathbf{X}_{\mathbf{i}};\boldsymbol{\beta}_{0}\right)
\overset{\text{def}}{=}\mathbf{D}\left(  \mathbf{X}_{\mathbf{i}}\right)
\text{,} \label{eqn.Dx.001.new1}%
\end{equation}
which is the partial derivatives of $\boldsymbol{\beta}$ for the mean function
$\mathbf{h}\left(  \mathbf{X}_{\mathbf{i}};\boldsymbol{\beta}_{0}\right)  $ in
(\ref{EQN.frm.01}). Then the efficient score in (\ref{eqn.n18}) simplifies to
\
\begin{equation}
\mathbf{S}_{\text{eff1}}\left(  \varepsilon_{\mathbf{i}},\mathbf{X}%
_{\mathbf{i}};\boldsymbol{\theta}_{0}\right)  =E\left[  \mathbf{S}%
_{\boldsymbol{\beta}}\left(  \varepsilon_{\mathbf{i}},\mathbf{X}_{\mathbf{i}%
}\right)  \varepsilon_{\mathbf{i}}\mid \mathbf{X}_{\mathbf{i}}\right]
V^{-1}\left(  \mathbf{X}_{\mathbf{i}}\right)  \varepsilon_{\mathbf{i}%
}=\mathbf{D}^{\top}\left(  \mathbf{X}_{\mathbf{i}}\right)  V^{-1}\left(
\mathbf{X}_{\mathbf{i}}\right)  \varepsilon_{\mathbf{i}}. \label{eqn.FRM.1}%
\end{equation}
By (\ref{Thm9}) in Theorem 7, the unique efficient I.F. for model class 1 is
obtained by scaling $\mathbf{S}_{\text{eff1}}\left(  \varepsilon_{\mathbf{i}%
},\mathbf{X}_{\mathbf{i}};\boldsymbol{\theta}_{0}\right)  $:
\begin{equation}
\boldsymbol{\varphi}_{\text{eff1}}\left(  \varepsilon_{\mathbf{i}}%
,\mathbf{X}_{\mathbf{i}};\boldsymbol{\theta}_{0}\right)  =E^{-1}\left(
\mathbf{S}_{\text{eff1}}\mathbf{S}_{\text{eff1}}^{\top}\right)  \mathbf{S}%
_{\text{eff1}}=E^{-1}\left(  \mathbf{D}_{\mathbf{i}}^{\top}V_{\mathbf{i}}%
^{-1}\mathbf{D}_{\mathbf{i}}\right)  \mathbf{D}_{\mathbf{i}}^{\top
}V_{\mathbf{i}}^{-1}\left[  f_{\mathbf{i}}-h\left(  \mathbf{X}_{\mathbf{i}%
};\boldsymbol{\beta}_{0}\right)  \right]  , \label{eqn.n20}%
\end{equation}
which is easily verified to satisfy (i) - (iii) in (\ref{eqn.semi}).

By Theorem 8, this semiparametric efficient I.F. $\boldsymbol{\varphi
}_{\text{eff1}}\left(  \varepsilon_{\mathbf{i}},\mathbf{X}_{\mathbf{i}%
};\boldsymbol{\theta}_{0}\right)  $ is in the equivalence class of the
efficient I.F. for model class 2, thus achieving the semiparametric efficiency
bound $\upsilon_{2}:$%
\begin{equation}
\upsilon_{2}=Var\left \{  \mathcal{M}\left[  \boldsymbol{\varphi}_{\text{eff1}%
}\left(  \varepsilon_{\mathbf{i}},\mathbf{X}_{\mathbf{i}};\boldsymbol{\theta
}_{0}\right)  \right]  \right \}  =Var\left[  2E\left(  \boldsymbol{\varphi
}_{\text{eff1}}\left(  \varepsilon_{\mathbf{i}},\mathbf{X}_{\mathbf{i}%
};\boldsymbol{\theta}_{0}\right)  \mid \mathbf{Z}_{i_{1}}\right)  \right]
=\mathbf{B}^{-1}\mathbf{\Sigma}_{U}\mathbf{B}^{-1}, \label{Bound}%
\end{equation}
where
\begin{align}
\mathbf{B}  &  =E\left[  \mathbf{D}^{\top}\left(  \mathbf{X}_{\mathbf{i}%
}\right)  V^{-1}\left(  \mathbf{X}_{\mathbf{i}}\right)  \mathbf{D}\left(
\mathbf{X}_{\mathbf{i}}\right)  \right]  ,\text{ }\widetilde{\mathbf{v}%
}_{i_{1}}=2E\left \{  \mathbf{D}^{\top}\left(  \mathbf{X}_{\mathbf{i}}\right)
V\left(  \mathbf{X}_{\mathbf{i}}\right)  ^{-1}\left[  f_{\mathbf{i}}%
-\mu \left(  \mathbf{X}_{\mathbf{i}},\boldsymbol{\beta}_{0}\right)  \right]
\mid \mathbf{Z}_{i_{1}}\right \}  ,\nonumber \\
\mathbf{\Sigma}_{U}  &  =Var\left(  \widetilde{\mathbf{v}}_{i_{1}}\right)
=\mathbf{v}_{i_{1}}\mathbf{v}_{i_{1}}^{\top},\text{ }\mathbf{i}=\left(
i_{1},i_{2}\right)  \in C_{2}^{n},\text{ }\mathbf{Z}_{i_{1}}=(\mathbf{Y}%
_{i_{1}}^{\top},\mathbf{X}_{i_{1}}^{\top})^{\top}. \label{eqn.n22}%
\end{align}
Consequently, the efficient score equations\
\begin{equation}
\sum_{\mathbf{i\in}C_{2}^{n}}\mathbf{S}_{\text{eff1}}\left(  \varepsilon
_{\mathbf{i}},\mathbf{X}_{\mathbf{i}}\right)  =\sum_{\mathbf{i\in}C_{2}^{n}%
}\mathbf{D}^{\top}(\mathbf{X}_{\mathbf{i}})V^{-1}(\mathbf{X}_{\mathbf{i}%
})\left[  f_{\mathbf{i}}-h\left(  \mathbf{X}_{\mathbf{i}},\boldsymbol{\beta
}\right)  \right]  =\mathbf{0,} \label{eqn.n23}%
\end{equation}
yield an estimator $\widehat{\boldsymbol{\beta}}_{\text{eff}}$ whose variance
(after mapping) is the smallest among all semiparametric RAL estimators of the FRM.

This $\upsilon_{2}$ coincides with $\mathbf{\Sigma}_{\boldsymbol{\beta}%
}^{\text{ugee}}$ in (\ref{eqn.thm1}), which is the asymptotic variance of the
UGEE estimator. Hence, the UGEE in (\ref{eqn140}) for between-subject FRM is
exactly the efficient estimating equation (\ref{eqn.n23}), and the resulting
UGEE estimator does achieve the semiparametric efficiency bound $\upsilon_{2}%
$, provided $V\left(  \mathbf{X}_{\mathbf{i}}\right)  $ is specified correctly.

\section{Examples of UGEE and Efficient I.F. \label{Ef.sec4.4}}

In this section, we demonstrate examples of UGEE\ estimators that achieve the
semiparametric efficiency bound. For space consideration, more examples of
binary or count responses are included in the Supplements.

\subsection{Exogenous\ Between-subject Responses\label{Ef.sec4.4.1}}

Consider a classical linear regression
\[
Y_{i}=X_{i}\beta+\varepsilon_{i},\text{ }Y_{i}\sim^{i.i.d}N\left(
0,\sigma_{Y}^{2}\right)  ,\text{ }1\leq i\leq n,
\]
assume $X_{i}\sim^{i.i.d}N\left(  0,\sigma_{X}^{2}\right)  $ without loss of
generality. The maximum likelihood estimator (MLE) of $\beta$ reaches the
Cram\'{e}r-Rao (CR) bound $\sigma_{Y}^{2}\sigma_{X}^{-2}.$

Now construct (exogenous) between-subject attributes for the $i$-th pair with
$f_{\mathbf{i}}=Y_{i_{1}}-Y_{i_{2}}$ and $X_{\mathbf{i}}=X_{i_{1}}-X_{i_{2}}$.
Consider an FRM that restricts the mean $E\left(  f_{\mathbf{i}}\mid
X_{\mathbf{i}}\right)  =X_{\mathbf{i}}\beta$. Let
\[
S_{\mathbf{i}}=f_{\mathbf{i}}-X_{\mathbf{i}}\beta,\text{ }D_{\mathbf{i}}%
=\frac{\partial}{\partial \beta}\left(  X_{\mathbf{i}}\beta \right)
=X_{\mathbf{i}},\text{ }V_{\mathbf{i}}=Var\left(  f_{\mathbf{i}}\right)
=2\sigma_{Y}^{2}.
\]
The UGEE and associated I.F. are given by
\begin{align}
U_{n}\left(  \beta \right)   &  =\sum_{\mathbf{i\in}C_{2}^{n}}D_{\mathbf{i}%
}V_{\mathbf{i}}^{-1}S_{\mathbf{i}}=\sum_{\mathbf{i\in}C_{2}^{n}}X_{\mathbf{i}%
}\left(  2\sigma_{Y}^{2}\right)  ^{-1}\left(  f_{\mathbf{i}}-X_{\mathbf{i}%
}\beta \right)  =0\mathbf{,}\text{ }\label{eg1}\\
\varphi_{\text{ugee}}\left(  \varepsilon_{\mathbf{i}},X_{\mathbf{i}};\beta
_{0}\right)   &  =E(X_{\mathbf{i}}V_{\mathbf{i}}^{-1}X_{\mathbf{i}%
})X_{\mathbf{i}}V_{\mathbf{i}}^{-1}\left(  f_{\mathbf{i}}-X_{\mathbf{i}}%
\beta_{0}\right)  =\left(  2\sigma_{X}^{2}\right)  ^{-1}\left(  \varepsilon
_{\mathbf{i}}X_{\mathbf{i}}\right)  .\nonumber
\end{align}
The asymptotic variance calculated based on norm b2 is
\[
\upsilon_{2}=\left \Vert \varphi_{\text{ugee}}\left(  \varepsilon_{\mathbf{i}%
},X_{\mathbf{i}};\beta_{0}\right)  \right \Vert _{b2}^{2}=Var\left[  2E\left(
\varphi_{\text{ugee}}\left(  \varepsilon_{\mathbf{i}},X_{\mathbf{i}};\beta
_{0}\right)  \mid \varepsilon_{i_{1}},X_{i_{1}}\right)  \right]  =\sigma
_{Y}^{2}\sigma_{X}^{-2},
\]
which exactly achieves the CR bound for the MLE of $\beta$ in the classic
linear regression. Hence, the semiparamatric UGEE estimator by solving for
(\ref{eg1}) is efficient.

\subsection{Endogenous\ Between-subject Responses\label{Ef.sec4.4.2}}

Now consider $i.d.$ (identically but not independently distributed)
endogenous\ between-subject responses $f_{\mathbf{i}}=f_{i_{1},i_{2}}$ with
mean $\mu$ and variance $\sigma^{2}$, where, unlike the exogenous\ example
above, their subject-level outcomes may be latent. Denote $\boldsymbol{\beta
}=\left(  \mu,\sigma^{2}\right)  ^{\top}$ the parameters of interest and
$\boldsymbol{\beta}_{0}=\left(  \mu_{0},\sigma_{0}^{2}\right)  ^{\top}$ the
truth. To first obtain an efficient (parametric) estimator for
$\boldsymbol{\beta}$ as our benchmark, we assume $f_{\mathbf{i}}\sim
^{i.d.}N\left(  \mu,\sigma^{2}\right)  $. Then the efficient (parametric) I.F.
for model class 1 is
\begin{equation}
\boldsymbol{\varphi}_{\text{eff1}}\left(  f_{\mathbf{i}}\right)  =\left(
f_{\mathbf{i}}-\mu_{0},\text{ }-\sigma_{0}^{2}+\left(  f_{\mathbf{i}}-\mu
_{0}\right)  ^{2}\right)  ^{\top}, \label{eqn.eg.2}%
\end{equation}
which is also in the equivalent class of\ the efficient I.F. $\boldsymbol{\psi
}_{\text{eff2}}\left(  f_{\mathbf{i}}\right)  $ for the model class 2 whose
variance (based on norm b2) is%
\[
\mathbf{\Sigma}_{\boldsymbol{\beta}_{0}}^{\text{eff2}}=4E\left[  E\left(
\boldsymbol{\varphi}_{\text{eff1}}\left(  f_{\mathbf{i}}\right)  \mid
f_{i_{1}}\right)  E\left(  \boldsymbol{\varphi}_{\text{eff1}}^{\top}\left(
f_{\mathbf{i}}\right)  \mid f_{i_{1}}\right)  \right]  .
\]
For endogenous\ responses where the benchmark based on individuals is
intractable, we use this $\mathbf{\Sigma}_{\boldsymbol{\beta}_{0}%
}^{\text{eff2}}$ as the efficiency bound.

Now consider a semiparametric FRM by modeling $E\left(  f_{\mathbf{i}}\right)
=\mu,$ $E$ $\left[  {\normalsize \left(  f_{\mathbf{i}}-\mu \right)  ^{2}{}%
}\right]  =\sigma^{2}$, let%
\begin{equation}
\mathbf{S}_{\mathbf{i}}=\left(  f_{\mathbf{i}}-\mu,\text{ }\left(
f_{\mathbf{i}}-\mu \right)  ^{2}-\sigma^{2}\right)  ^{\top},\text{ }%
\mathbf{D}_{\mathbf{i}}=\frac{\partial}{\partial \boldsymbol{\beta}^{\top}%
}\boldsymbol{\beta},\text{ }\mathbf{V}_{\mathbf{i}}=diag\left(  Var\left(
f_{\mathbf{i}}\right)  ,\text{ }Var\left[  \left(  f_{\mathbf{i}}-\mu \right)
^{2}\right]  \right)  , \label{eqn.eg.n7}%
\end{equation}
The UGEE, the resulting estimator, and the associated I.F. for the FRM are
\begin{align*}
\mathbf{U}_{n}\left(  \boldsymbol{\beta}\right)   &  =\sum_{\mathbf{i\in}%
C_{2}^{n}}\mathbf{D}_{\mathbf{i}}^{\top}\mathbf{V}_{\mathbf{i}}^{-1}%
\mathbf{S}_{\mathbf{i}}=\mathbf{0},\text{ }\widehat{\boldsymbol{\beta}}%
_{f}^{\text{ugee}}=\binom{n}{2}^{-1}\sum_{\mathbf{i}\in C_{2}^{n}}\left(
f_{\mathbf{i}},\text{ }\left(  f_{\mathbf{i}}-\overline{f_{\mathbf{i}}%
}\right)  ^{2}\right)  ^{\top},\\
\boldsymbol{\varphi}_{\text{ugee}}\left(  f_{\mathbf{i}}\right)   &  =\left(
f_{\mathbf{i}}-\mu_{0},\text{ }-\sigma_{0}^{2}+\left(  f_{\mathbf{i}}-\mu
_{0}\right)  ^{2}\right)  ^{\top}.
\end{align*}
Since here $\boldsymbol{\varphi}_{\text{ugee}}\left(  f_{\mathbf{i}}\right)
=\boldsymbol{\varphi}_{\text{eff1}}\left(  f_{\mathbf{i}}\right)  $ in
(\ref{eqn.eg.2}), this $\widehat{\boldsymbol{\beta}}_{f}^{\text{ugee}}$does
achieve the benchmark $\mathbf{\Sigma}_{\boldsymbol{\beta}_{0}}^{\text{eff2}}%
$\ and hence is optimal. Therefore, in the endogenous case, UGEE also yields
the most efficient semiparametric RAL estimator.

\section{Adaptive Semiparametric Estimator for FRM\label{Ef.sec7}}

Recall that we have proved that UGEE estimator achieves the semiparametric
efficiency bound $\upsilon_{2}$, provided $V\left(  \mathbf{X}_{\mathbf{i}%
}\right)  $ is specified correctly. Here we need to differentiatelocal and
global efficiency. Local efficiency refers to the efficiency for particular
assumptions of the nonparametric component of the model. Such estimators are
optimal for a particular distribution, subject to the constraint implied by
the semiparametric model \cite{tsiatis2004}$,$ while the more ambitious global
efficiency refers to the efficiency for all values of the nonparametric
component \cite{Bickle1998}$.$

We define local and global efficiency for FRM in the same vein as for
within-subject models. Namely, any semiparametric RAL estimator $\widehat
{\boldsymbol{\beta}}$ with the asymptotic variance achieving the bound
$\upsilon_{2}$ in (\ref{Bound}) for the true model $p_{0}(f_{\mathbf{i}%
},\mathbf{X}_{\mathbf{i}})=p\left(  f_{\mathbf{i}},\mathbf{X}_{\mathbf{i}%
};\boldsymbol{\theta}_{0}\right)  $ is locally efficient\ at\ $p_{0}\left(
f_{\mathbf{i}},\mathbf{X}_{\mathbf{i}}\right)  $. If the same $\widehat
{\boldsymbol{\beta}}$ is semiparametric efficient regardless of $p_{0}%
(f_{\mathbf{i}},\mathbf{X}_{\mathbf{i}})\in \mathcal{P}$, then it is globally
efficient.\ For FRM, the nonparametric component refers to the unknown true
conditional distribution $p_{0}\left(  f_{\mathbf{i}}\mid \mathbf{X}%
_{\mathbf{i}}\right)  $ left unspecified, which yields an unknown conditional
variance $V\left(  \mathbf{X}_{\mathbf{i}}\right)  =Var\left(  f_{\mathbf{i}%
}\mid \mathbf{X}_{\mathbf{i}}\right)  .$

To resolve this chicken and egg situation, a feasible approach is adaptive
estimators \cite{Tsaitis2006}$,$ where we find approximations to $V\left(
\mathbf{X}_{\mathbf{i}}\right)  $ by imposing additional assumptions to
improve efficiency, as shown in our simulations (see Section \ref{Ef.sec7.3}
and Supplements). In the following, we discuss global and local efficiency for FRM.

\subsection{Globally Efficient Estimators\label{Ef.sec7.1}}

\textbf{Example 1. (Binary responses)} \ Consider an FRM for binary responses
$f_{\mathbf{i}}$ with a vector of explanatory variables $\mathbf{X}%
_{\mathbf{i}}$, where $E\left(  f_{\mathbf{i}}\mid \mathbf{X}_{\mathbf{i}%
}\right)  =$ expit$\left(  \boldsymbol{\beta}^{\top}\mathbf{X}_{\mathbf{i}%
}\right)  =\exp \left(  \boldsymbol{\beta}^{\top}\mathbf{X}_{\mathbf{i}%
}\right)  \left[  1+\exp \left(  \boldsymbol{\beta}^{\top}\mathbf{X}%
_{\mathbf{i}}\right)  \right]  ^{-1}$. The variance of the binary
$f_{\mathbf{i}}$ conditional on $\mathbf{X}_{\mathbf{i}}$ takes the form%
\begin{equation}
V\left(  \mathbf{X}_{\mathbf{i}};\boldsymbol{\beta}\right)  =\exp \left(
\boldsymbol{\beta}^{\top}\mathbf{X}_{\mathbf{i}}\right)  \left[  1+\exp \left(
\boldsymbol{\beta}^{\top}\mathbf{X}_{\mathbf{i}}\right)  \right]  ^{-2},
\label{eqn.202}%
\end{equation}
which does not involve any additional unknown parameter (aside from $\beta$).
By (\ref{eqn.n20}), the optimal UGEE is:%
\[
\sum_{\mathbf{i\in}C_{2}^{n}}\mathbf{D}_{\mathbf{i}}^{\top}V_{\mathbf{i}}%
^{-1}S_{\mathbf{i}}=\sum_{\mathbf{i\in}C_{2}^{n}}\mathbf{X}_{\mathbf{i}%
}{\left[  f_{\mathbf{i}}-\text{expit}\left(  \boldsymbol{\beta}^{\top
}\mathbf{X}_{\mathbf{i}}\right)  \right]  =}\text{ }\mathbf{0.}%
\]
Since the above only contains $\boldsymbol{\beta}$ with no other parameter,
the resulting UGEE estimator $\widehat{\boldsymbol{\beta}}$ has the\ efficient
I.F. depending only on $\boldsymbol{\beta}_{0}:$
\begin{equation}
\boldsymbol{\varphi}_{\text{eff1}}\left(  f_{\mathbf{i}},\mathbf{X}%
_{\mathbf{i}};\boldsymbol{\beta}_{0}\right)  =E^{-1}\left[  \mathbf{X}%
_{\mathbf{i}}V(\mathbf{X}_{\mathbf{i}};\boldsymbol{\beta}_{0})\mathbf{X}%
_{\mathbf{i}}^{\top}\right]  \mathbf{X}_{\mathbf{i}}\left[  f_{\mathbf{i}%
}-\text{expit}\left(  \boldsymbol{\beta}_{0}^{\top}\mathbf{X}_{\mathbf{i}%
}\right)  \right]  .\nonumber
\end{equation}
This $\widehat{\boldsymbol{\beta}}$ is semiparametric efficient regardless of
$p\left(  f_{\mathbf{i}},\mathbf{X}_{\mathbf{i}};\boldsymbol{\theta}%
_{0}\right)  \in \mathcal{P}$ and thus is globally efficient. \ 

\subsection{Locally Efficient Estimators\label{Ef.sec7.2}}

\textbf{Example 2. (Count responses)} \ Consider modeling count responses
$f_{\mathbf{i}}$ with $E\left(  f_{\mathbf{i}}\mid \mathbf{X}_{\mathbf{i}%
}\right)  =\exp \left(  \boldsymbol{\beta}^{\top}\mathbf{X}_{\mathbf{i}%
}\right)  $, where $f_{\mathbf{i}}$\ is over-dispersed. For the unknown
$V\left(  \mathbf{X}_{\mathbf{i}}\right)  ,$ we can specify a working variance
that is proportional to the conditional mean, $V\left(  \mathbf{X}%
_{\mathbf{i}};\tau^{2},\boldsymbol{\beta}\right)  =\tau^{2}\exp \left(
\boldsymbol{\beta}^{\top}\mathbf{X}_{\mathbf{i}}\right)  $, with $\tau^{2}=1$
for non-overdispersed and $\tau^{2}>1$ for overdispersed $f_{\mathbf{i}}$. We
then estimate $\tau^{2}$ and $\beta$ by iterating between (1) minimizing the
squared sum of residuals $\left \{  \left[  f_{\mathbf{i}}-\exp \left(
\boldsymbol{\beta}^{\top}\mathbf{X}_{\mathbf{i}}\right)  \right]
^{2}-V\left(  \mathbf{X}_{\mathbf{i}};\tau^{2},\boldsymbol{\beta}\right)
\right \}  ^{2}$ for $\tau^{2}$ with a given $\widehat{\boldsymbol{\beta}}$ and
(2) solving the UGEE for $\beta$ with a given $\widehat{\tau}^{2}$, until convergence.

Under mild regularity conditions, $\widehat{\tau}^{2}\rightarrow_{p}\tau
_{\ast}^{2}$ (a constant may or may not be the truth), leading to a UGEE
estimator $\widehat{\boldsymbol{\beta}}^{P}$ with the efficient I.F.
\begin{equation}
\boldsymbol{\varphi}_{\text{eff1}}\left(  f_{\mathbf{i}},\mathbf{X}%
_{\mathbf{i}};\tau_{\ast}^{2},\boldsymbol{\beta}_{0}\right)  =E^{-1}\left[
\mathbf{X}_{\mathbf{i}}\exp \left(  \boldsymbol{\beta}_{0}^{\top}%
\mathbf{X}_{\mathbf{i}}\right)  \mathbf{X}_{\mathbf{i}}^{\top}\right]
\mathbf{X}_{\mathbf{i}}\left[  f_{\mathbf{i}}-\exp \left(  \boldsymbol{\beta
}_{0}^{\top}\mathbf{X}_{\mathbf{i}}\right)  \right]  . \label{eqn.206.n}%
\end{equation}
This estimator is locally efficient; if the conditional variance is indeed
proportional to the conditional mean, i.e., $\tau_{\ast}^{2}=\tau_{0}^{2}$,
then it is semiparametric efficient.

Alternatively, we can specify a working variance motivated by the form of
Negative Binomial (NB) distribution. With a dispersion parameter $\zeta$, we
substitute $\exp \left(  \boldsymbol{\beta}^{\top}\mathbf{X}_{\mathbf{i}%
}\right)  \left[  1+\zeta \exp \left(  \boldsymbol{\beta}^{\top}\mathbf{X}%
_{\mathbf{i}}\right)  \right]  $ in place of $V\left(  \mathbf{X}_{\mathbf{i}%
};\zeta,\boldsymbol{\beta}\right)  $, yielding an UGEE\ estimator
$\widehat{\boldsymbol{\beta}}^{NB}$ with a different I.F.
\begin{equation}
E^{-1}\left \{  \mathbf{X}_{\mathbf{i}}\left[  1+\zeta_{\ast}\exp \left(
\boldsymbol{\beta}_{0}^{\top}\mathbf{X}_{\mathbf{i}}\right)  \right]
^{-1}\exp \left(  \boldsymbol{\beta}_{0}^{\top}\mathbf{X}_{\mathbf{i}}\right)
\mathbf{X}_{\mathbf{i}}^{\top}\right \}  \mathbf{X}_{\mathbf{i}}\left[
1+\zeta_{\ast}\exp \left(  \boldsymbol{\beta}_{0}^{\top}\mathbf{X}_{\mathbf{i}%
}\right)  \right]  ^{-1}\left[  f_{\mathbf{i}}-\exp \left(  \boldsymbol{\beta
}_{0}^{\top}\mathbf{X}_{\mathbf{i}}\right)  \right]  . \label{eqn.2061}%
\end{equation}
Again, it has the form of the efficient I.F., but with respect to the limiting
point $\zeta_{\ast}$\ that may or may not be true. If the working variance is
the same as the true variance, then the resulting $\widehat{\boldsymbol{\beta
}}^{NB}$ is semiparametric efficient.

The distinct forms of efficient I.F.s between (\ref{eqn.206.n}) and
(\ref{eqn.2061}) result from different working variance assumptions we made.
For count responses, other forms of non-negative working variance can be
assumed, each leads to a different variance (or local efficiency bound) of
$\widehat{\boldsymbol{\beta}}$.

Adaptive estimators have been shown empirically to improve efficiency for
classical semiparametric GLMs for within-subject attributes \cite{Tsaitis2006}%
$.$ Our simulation studies also demonstrate this feature, some of which are
discussed below. \ 

\subsection{Simulation Studies\label{Ef.sec7.3}}

To illustrate the local efficiency of adaptive estimators, we consider again
overdispersed count responses. We generated data from the Negative Binomial
distribution and estimated parameters using both parametric and semiparametric
models (but with different working variances). For Monte Carlo (MC)
simulations, we set total MC\ iterations\ $M=1,000$ and sample sizes $n=100$,
$300$, $500$. All analyses are performed with the R software platform
\cite{R2010}$,$ with code optimized using Rcpp \cite{eddelbuettel2011rcpp} for
run-time improvement. We demonstrate between-subject attributes here, similar
performances of within-subject attributes are observed but omitted here. \ 

Without loss of generality, we include one continuous predictor. By first
generating $X_{i}\sim^{i.i.d}U\left(  a,b\right)  $ with $U\left(  a,b\right)
$ denoting a uniform distribution over $\left(  a,b\right)  $, we create
between-subject $X_{\mathbf{i}}$ with $X_{\mathbf{i}}=X_{i_{1}}+X_{i_{2}}$.
Given $X_{\mathbf{i}}$, we then generate $f_{\mathbf{i}}\sim NB\left(
\zeta,h\left(  X_{\mathbf{i}};\boldsymbol{\beta}\right)  \right)  $, where
$h(X_{\mathbf{i}};\boldsymbol{\beta})=\exp \left(  \beta_{0}+\beta
_{1}X_{\mathbf{i}}\right)  $ and $NB\left(  \zeta,\mu \right)  $ denotes a
Negative Binomial with mean $\mu$ and dispersion parameter $\zeta$.

We estimate $\boldsymbol{\beta}=$ $(\beta_{0},\beta_{1})^{\top}$ using (i) MLE
from Negative Binomial (NB); and (ii) semiparametric UGEE with working
variances from (1) NB, (2) Poisson and (3) as a constant (See the Supplement
S2 for details). We set $\zeta=10,$ $\beta_{0}=3,$ $\beta_{1}=3,$ $a=0,$ $b=1$
and report the parameter estimators (Est.), asymptotic (Asy.) and empirical
(Emp.) variances under different sample sizes in Table 1.

******************** Table 1 goes here ********************

The MLE from NB is the benchmark for efficiency in this setting. As expected,
Table 1 shows that UGEE estimators with the working variance of NB reach the
local efficiency bound, while the other two yield larger variances. As
expected, the constant working variance yields the largest variance, since the
Poisson working variance has a better approximation to the true variance than
a constant. Thus, akin to within-subject attributes, adaptive estimators
demonstrate efficiency gains for semiparametric models of between-subject
attributes as well, but the improvement depends on how well the working
variance resembles the true variance. \ 

\section{Discussion\label{Ef.sec8 copy(1)}}

By leveraging the Hilbert-space-based semiparametric efficiency theory, we
demonstrated that UGEE estimators are semiparametric efficient for functional
response models (FRM) based on between-subject attributes. Such estimators
deliver the smallest asymptotic variances among a class of regular and
asymptotic linear (RAL) estimators for this emerging class of semiparametric
models. Specifying mathematical distributions such as normality for
between-subject attributes is far more challenging than for their
within-subject counterparts, because between-subject attributes are not only
correlated, but generally follow more complex distributions. Extending the
semiparametric efficiency theories to between-subject attributes will not only
enrich the body of research on this topic, but will also greatly facilitate
the implementations of FRM\ for valid and efficient inference in practice. \ 

To show the efficiency of UGEE\ estimators for FRM, or model class 2, we first
generalized relevant results to between-subject attributes, such as asymptotic
linearity, regular estimators, and efficiency bounds. Since directly
establishing the efficiency theory is difficult for UGEE estimators, we also
introduced a class of models involving only a subset of independent pairs of
between-subject responses, or model class 1. Although this \textquotedblleft
conjugate\textquotedblright \ class of models has no practical utility given
its lower efficiency, this powerful tool helps determine the efficiency of
UGEE estimator for the FRM. By connecting estimators from the two classes of
models with a dual orthogonality property between their respective nuisance
tangent spaces, we pinpointed the efficient estimator for FRM through first
finding the efficient estimator for the \textquotedblleft
conjugate\textquotedblright \ model class. This is more straightforward by
leveraging the existing Hilbert-space-based semiparametric efficiency theory. \ 

Therefore, not only does UGEE enjoy the semiparametric robustness, but also
the efficiency in inference, just like its counterpart GEE for the classical
within-subject attributes. With blooming implementations of between-subject
attributes as effective summary metrics of high-dimensional data, our
developed efficiency will greatly propel applying FRM for scientific
discovery. \ 

One limitation is that we only focus on the efficiency bound for
semiparametric FRM when applied to the cross-sectional data. Extending the
results to clustered data such as repeated assessments in longitudinal studies
is the next goal to undertake, where major challenges are to address the
missing data arising from study dropouts and elucidate its impact on
estimators through different missing data mechanisms.

\begin{center}
\newpage

\textbf{Table 1}%

\begin{tabular}
[c]{cccccccc}\hline
Method & Assumption & \multicolumn{3}{c}{$\beta_{0}$} &
\multicolumn{3}{c}{$\beta_{1}$}\\ \hline
&  & \multicolumn{6}{c}{$n=100$}\\ \cline{3-8}
&  & Est. & \multicolumn{2}{c}{Variance} & Est. & \multicolumn{2}{c}{Variance}%
\\
&  &  & \multicolumn{2}{c}{Asy.} &  & \multicolumn{2}{c}{Asy.}\\
Working-MLE & NB & 2.9911 & \multicolumn{2}{c}{0.0002} & 2.9990 &
\multicolumn{2}{c}{0.0001}\\ \cline{3-8}
&  &  & Asy. & Emp. &  & Asy. & Emp.\\
UGEE & NB & 3.0000 & 0.0002 & 0.0002 & 3.0000 & 0.0001 & 0.0001\\
& Pois & 3.0003 & 0.0007 & 0.0007 & 2.9997 & 0.0005 & 0.0005\\
& Const. & 2.9989 & 0.0060 & 0.0062 & 3.0005 & 0.0028 & 0.0029\\ \hline
&  & \multicolumn{6}{c}{$n=300$}\\ \cline{3-8}
&  & Est. & \multicolumn{2}{c}{Variance} & Est. & \multicolumn{2}{c}{Variance}%
\\
Working-MLE & NB & 2.9970 & \multicolumn{2}{c}{1.79e-05} & 2.9997 &
\multicolumn{2}{c}{1.49e-05}\\ \cline{3-8}
&  &  & Asy. & Emp. &  & Asy. & Emp.\\
UGEE & NB & 2.9998 & 1.77e-05 & 1.82e-05 & 3.0002 & 1.47e-05 & 1.46e-05\\
& Pois & 3.0000 & 7.53e-05 & 7.36e-05 & 3.0000 & 5.12e-05 & 4.98e-05\\
& Const. & 3.0003 & 0.0007 & 0.0007 & 2.9998 & 0.0003 & 0.0003\\ \hline
&  & \multicolumn{6}{c}{$n=500$}\\ \cline{3-8}
&  & Est. & \multicolumn{2}{c}{Variance} & Est. & \multicolumn{2}{c}{Variance}%
\\
Working-MLE & NB & 2.9983 & \multicolumn{2}{c}{6.33e-06} & 2.9997 &
\multicolumn{2}{c}{5.28e-06}\\ \cline{3-8}
&  &  & Asy. & Emp. &  & Asy. & Emp.\\
UGEE & NB & 3.0000 & 6.27e-06 & 6.21e-06 & 3.0000 & 5.23e-06 & 5.08e-06\\
& Pois & 2.9998 & 2.73e-05 & 2.65e-05 & 3.0001 & 1.85e-05 & 1.78e-05\\
& Const. & 2.9998 & 0.0002 & 0.0002 & 3.0001 & 0.0001 & 0.0001\\ \hline
\end{tabular}

\end{center}

\bigskip \newpage

\begin{center}
\bigskip \textbf{Supplemental Material}
\end{center}

\textbf{S1. Details about the Hilbert Space for Between-subject Attributes}
\label{Ef.A copy(1)}

For the \textit{norm b2} of the between-subject attributes that encompass an
FRM form for the correlated\emph{ }$\mathbf{\mathbf{h}}\left(  \mathbf{Z}%
_{\mathbf{i}}\right)  $'s, we equipped the Hilbert space $\mathcal{H}%
_{b}^{(q)}$ with%

\begin{align*}
\left \langle \mathbf{h}_{1}\left(  \mathbf{Z}_{\mathbf{i}}\right)
,\mathbf{h}_{2}\left(  \mathbf{Z}_{\mathbf{i}}\right)  \right \rangle _{b2}  &
=E\left \{  2E\left[  \mathbf{h}_{1}^{\top}\left(  \mathbf{Z}_{\mathbf{i}%
}\right)  \mid \mathbf{Z}_{i_{1}}\right]  \cdot2E\left[  \mathbf{h}_{2}\left(
\mathbf{Z}_{\mathbf{i}}\right)  \mid \mathbf{Z}_{i_{1}}\right]  \right \}  ,\\
\left \Vert \mathbf{h}\left(  \mathbf{Z}_{\mathbf{i}}\right)  \right \Vert
_{b2}  &  =\left \langle \mathbf{h}\left(  \mathbf{Z}_{\mathbf{i}}\right)
,\mathbf{h}\left(  \mathbf{Z}_{\mathbf{i}}\right)  \right \rangle _{b2}%
^{1/2}=E^{1/2}\left \{  2E\left[  \mathbf{h}^{\top}\left(  \mathbf{Z}%
_{\mathbf{i}}\right)  \mid \mathbf{Z}_{i_{1}}\right]  \cdot2E\left[
\mathbf{h}\left(  \mathbf{Z}_{\mathbf{i}}\right)  \mid \mathbf{Z}_{i_{1}%
}\right]  \right \}  .
\end{align*}

It is readily checked that this definition of inner product 2 satisfies
conditions 1) - 3) below,
\begin{align*}
1).\text{ }\left \langle \mathbf{h}_{1}\left(  \mathbf{Z}_{\mathbf{i}}\right)
,\mathbf{h}_{2}\left(  \mathbf{Z}_{\mathbf{i}}\right)  \right \rangle _{b2}  &
=\left \langle \mathbf{h}_{2}\left(  \mathbf{Z}_{\mathbf{i}}\right)
,\mathbf{h}_{1}\left(  \mathbf{Z}_{\mathbf{i}}\right)  \right \rangle _{b2},\\
2).\text{ }\left \langle a\mathbf{h}_{1}\left(  \mathbf{Z}_{\mathbf{i}}\right)
,\mathbf{h}_{2}\left(  \mathbf{Z}_{\mathbf{i}}\right)  \right \rangle _{b2}  &
=a\left \langle \mathbf{h}_{1}\left(  \mathbf{Z}_{\mathbf{i}}\right)
,\mathbf{h}_{2}\left(  \mathbf{Z}_{\mathbf{i}}\right)  \right \rangle _{b2},\\
3).\text{ }\left \langle \mathbf{h}_{1}\left(  \mathbf{Z}_{\mathbf{i}}\right)
+\mathbf{h}_{2}\left(  \mathbf{Z}_{\mathbf{i}}\right)  ,\mathbf{h}_{3}\left(
\mathbf{Z}_{\mathbf{i}}\right)  \right \rangle _{b2}  &  =\left \langle
\mathbf{h}_{1}\left(  \mathbf{Z}_{\mathbf{i}}\right)  ,\mathbf{h}_{2}\left(
\mathbf{Z}_{\mathbf{i}}\right)  \right \rangle _{b2}+\left \langle
\mathbf{h}_{1}\left(  \mathbf{Z}_{\mathbf{i}}\right)  ,\mathbf{h}_{3}\left(
\mathbf{Z}_{\mathbf{i}}\right)  \right \rangle _{b2},\\
4).\text{ }\left \langle \mathbf{h}\left(  \mathbf{Z}_{\mathbf{i}}\right)
,\mathbf{h}\left(  \mathbf{Z}_{\mathbf{i}}\right)  \right \rangle _{b2}  &
\geq0,\text{ }\left \langle \mathbf{h}\left(  \mathbf{Z}_{\mathbf{i}}\right)
,\mathbf{h}\left(  \mathbf{Z}_{\mathbf{i}}\right)  \right \rangle _{b2}=0\text{
iff }E\left[  \mathbf{h}\left(  \mathbf{Z}_{\mathbf{i}}\right)  \mid
\mathbf{Z}_{i_{1}}\right]  =\mathbf{0}\text{ a.s..}%
\end{align*}
For 4), we have that if $E\left[  \mathbf{h}\left(  \mathbf{Z}_{\mathbf{i}%
}\right)  \mid \mathbf{Z}_{i_{1}}\right]  =\mathbf{0}$ a.s., then $\left \Vert
\mathbf{h}\left(  \mathbf{Z}_{\mathbf{i}}\right)  \right \Vert _{b2}%
^{2}=\left \langle \mathbf{h}\left(  \mathbf{Z}_{\mathbf{i}}\right)
,\mathbf{h}\left(  \mathbf{Z}_{\mathbf{i}}\right)  \right \rangle _{b2}=0$.
Conversely, $\left \langle \mathbf{h}\left(  \mathbf{Z}_{\mathbf{i}}\right)
,\mathbf{h}\left(  \mathbf{Z}_{\mathbf{i}}\right)  \right \rangle _{b2}=0$
implies that for all $1\leq s\leq q,$
\[
E\left \{  E\left[  h_{s}\left(  \mathbf{Z}_{\mathbf{i}}\right)  \mid
\mathbf{Z}_{i_{1}}\right]  E\left[  h_{s}\left(  \mathbf{Z}_{\mathbf{i}%
}\right)  \mid \mathbf{Z}_{i_{1}}\right]  \right \}  =E\left \{  E^{2}\left[
h_{s}\left(  \mathbf{Z}_{\mathbf{i}}\right)  \mid \mathbf{Z}_{\mathbf{i}i_{1}%
}\right]  \right \}  =0,
\]
we then have:
\[
E\left[  h_{s}\left(  \mathbf{Z}_{\mathbf{i}}\right)  \mid \mathbf{Z}_{i_{1}%
}\right]  =0\text{ }a.s.\text{ for all }1\leq s\leq q,\text{ i.e., }E\left[
\mathbf{h}\left(  \mathbf{Z}_{\mathbf{i}}\right)  \mid \mathbf{Z}_{i_{1}%
}\right]  =\mathbf{0}\text{ }a.s.\text{.}%
\]

Thus,
\[
\left \langle \mathbf{h}\left(  \mathbf{Z}_{\mathbf{i}}\right)  ,\mathbf{h}%
\left(  \mathbf{Z}_{\mathbf{i}}\right)  \right \rangle _{b2}=0\text{ iff
}E\left[  \mathbf{h}\left(  \mathbf{Z}_{\mathbf{i}}\right)  \mid
\mathbf{Z}_{i_{1}}\right]  =\mathbf{0}\text{ }a.s..
\]

In general, $\left \langle \mathbf{h}\left(  \mathbf{Z}_{\mathbf{i}}\right)
,\mathbf{h}\left(  \mathbf{Z}_{\mathbf{i}}\right)  \right \rangle _{b2}=0$ does
not imply $\mathbf{h}\left(  \mathbf{Z}_{\mathbf{i}}\right)  =\mathbf{0}$
$a.s.$.\ To see this, consider a counterexample
\[
Z_{i_{1}},Z_{i_{2}}\sim N\left(  1,1\right)  ,\quad h\left(  Z_{\mathbf{i}%
}\right)  =h\left(  Z_{i_{1}},Z_{i_{2}}\right)  =\left(  1-Z_{i_{1}}\right)
\left(  1-Z_{i_{2}}\right)  .
\]
Then, $h\left(  Z_{\mathbf{i}}\right)  =h\left(  Z_{i_{1}},Z_{i_{2}}\right)  $
is summetric and although $\left \langle h\left(  Z_{\mathbf{i}}\right)
,h\left(  Z_{\mathbf{i}}\right)  \right \rangle _{b2}=\left \Vert h\left(
Z_{\mathbf{i}}\right)  \right \Vert _{b2}^{2}=0$, since%

\[
E\left[  h\left(  Z_{\mathbf{i}}\right)  \mid Z_{i_{1}}\right]  =\left(
1-Z_{i_{1}}\right)  E\left(  1-Z_{i_{2}}\right)  =0\text{ }a.s.,
\]
in general,
\[
h\left(  Z_{\mathbf{i}}\right)  \neq0\text{ }a.s.,
\]
i.e., here $\left \Vert h\left(  Z_{\mathbf{i}}\right)  \right \Vert _{b2}%
^{2}=0$ iff $E\left[  h\left(  Z_{\mathbf{i}}\right)  \mid Z_{i_{1}}\right]
=0$ a.s., but $\left \Vert h\left(  Z_{\mathbf{i}}\right)  \right \Vert
_{b2}^{2}=0$ does not imply $h\left(  Z_{\mathbf{i}}\right)  =0$ a.s..

Thus, unlike the origin of $\mathcal{H}_{b}$ under the inner product 1, the
origin\emph{ }of $\mathcal{H}_{b}$ under inner product 2 is not the
equivalence class of $\mathbf{h}\left(  \mathbf{Z}_{\mathbf{i}}\right)  $ with
$\mathbf{h}\left(  \mathbf{Z}_{\mathbf{i}}\right)  =\mathbf{0}$ a.s., but a
larger equivalence class consisting of functions $\mathbf{h}\left(
\mathbf{Z}_{\mathbf{i}}\right)  $ such that $E\left[  \mathbf{h}\left(
\mathbf{Z}_{\mathbf{i}}\right)  \mid \mathbf{Z}_{i_{1}}\right]  =\mathbf{0}$ a.s..

\textbf{S2. Detailed Simulation Settings}\label{Ef.A2}

We conduct a similar simulation for between-subject attributes, to demonstrate
the local efficiency of UGEE for count responses. We first simulate $X_{i}%
\sim^{i.i.d}Unif(a,b),$ then\ construct $X_{\mathbf{i}}=X_{i_{1}}+X_{i_{2}}.$ Let%

\[
E(f_{\mathbf{i}}\mid x_{\mathbf{i}})=\exp(\beta_{0}+\beta_{1}x_{\mathbf{i}%
})=h_{\mathbf{i}}(\boldsymbol{\beta}),\text{ }\boldsymbol{\beta}=(\beta
_{0},\beta_{1}),
\]
we can simulate overdispersed $f_{\mathbf{i}}\sim NB(\tau,h_{\mathbf{i}%
}(\boldsymbol{\beta}))$ following a Negative Binomial distribution with mean
$h_{\mathbf{i}}(\boldsymbol{\beta})$ and dispersion paramater $\tau$ (or the
shape parameter of the gamma mixing distribution)$.$

We then estimate $\boldsymbol{\beta}$ using

1) The working-MLE of Negative Binomial through\textbf{ }$f_{\mathbf{i}};$

2) Semiparametric UGEE with\textbf{ }%

\[
\mathbf{U}_{n}\left(  \boldsymbol{\beta}\right)  =\sum_{\mathbf{i\in}C_{2}%
^{n}}\mathbf{D}_{\mathbf{i}}^{\top}V_{\mathbf{i}}^{-1}S_{\mathbf{i}%
},\ S_{\mathbf{i}}=f_{\mathbf{i}}-h_{\mathbf{i}},\  \mathbf{D}_{\mathbf{i}%
}=\frac{\partial}{\partial \boldsymbol{\beta}^{\top}}h_{\mathbf{i}%
}(\boldsymbol{\beta}).
\]

For the unknown $V_{\mathbf{i}}$, we respectively chose

\  \ a) the true variance of Negative Binomial for\textbf{ }$f_{\mathbf{i}}.$ Let%

\[
V_{\mathbf{i}}=Var\left(  f_{\mathbf{i}}\mid x_{\mathbf{i}}\right)
=\frac{h_{\mathbf{i}}(\boldsymbol{\beta})}{p_{\mathbf{i}}(\boldsymbol{\beta}%
)},\quad p_{\mathbf{i}}(\boldsymbol{\beta})=\frac{\tau}{\tau+h_{\mathbf{i}%
}(\boldsymbol{\beta})}.
\]
The optimal UGEE for estimating $\boldsymbol{\beta}$ then becomes
\begin{equation}
\mathbf{U}_{n}\left(  \boldsymbol{\beta}\right)  =\sum_{\mathbf{i\in}C_{2}%
^{n}}\mathbf{D}_{\mathbf{i}}^{\top}V_{\mathbf{i}}^{-1}S_{\mathbf{i}}%
=\sum_{\mathbf{i\in}C_{2}^{n}}\mathbf{X}_{\mathbf{i}}^{\top}p_{\mathbf{i}%
}(\boldsymbol{\beta})\left[  f_{\mathbf{i}}-h_{\mathbf{i}}(\boldsymbol{\beta
})\right]  =0,
\end{equation}
yielding the asymptotic variance of%

\begin{align*}
Var(\boldsymbol{\beta})  &  =\mathbf{B}^{-1}4Var\left[  \mathbf{X}%
_{\mathbf{i}}^{\top}p_{\mathbf{i}}(\boldsymbol{\beta})\left \{  f_{\mathbf{i}%
}-h_{\mathbf{i}}(\boldsymbol{\beta})\right \}  \mid f_{i_{1}},\mathbf{X}%
_{i_{1}}\right]  \mathbf{B}^{-1},\\
\mathbf{B}  &  =E\left[  \mathbf{X}_{\mathbf{i}}^{\top}p_{\mathbf{i}%
}(\boldsymbol{\beta})h_{\mathbf{i}}(\boldsymbol{\beta})E\left \{
h_{\mathbf{i}}(\boldsymbol{\beta})\right \}  \right]  .
\end{align*}
In the simulation, we estimate $\tau$ from the sample$.$

\ b) a (wrong) variance assumption of\textbf{ }$Poisson$\textbf{ }for\textbf{
}$\theta$\textbf{\ }through\textbf{ }$f_{\mathbf{i}}.$ Let%

\[
V_{\mathbf{i}}=Var\left(  f_{\mathbf{i}}\mid x_{\mathbf{i}}\right)
=h_{\mathbf{i}}(\boldsymbol{\beta}).
\]
The optimal UGEE for estimating $\boldsymbol{\beta}$ is%

\begin{equation}
\mathbf{U}_{n}\left(  \boldsymbol{\beta}\right)  =\sum_{\mathbf{i\in}C_{2}%
^{n}}\mathbf{D}_{\mathbf{i}}^{\top}V_{\mathbf{i}}^{-1}S_{\mathbf{i}}%
=\sum_{\mathbf{i\in}C_{2}^{n}}\mathbf{X}_{\mathbf{i}}^{\top}\left \{
f_{\mathbf{i}}-h_{\mathbf{i}}(\boldsymbol{\beta})\right \}  =0,
\end{equation}
yielding the asymptotic variance of%

\[
Var(\boldsymbol{\beta})=E\left[  \mathbf{X}_{\mathbf{i}}^{\top}h_{\mathbf{i}%
}(\boldsymbol{\beta})\mathbf{X}_{\mathbf{i}}\right]  ^{-1}4Var\left[
\mathbf{X}_{\mathbf{i}}^{\top}\left \{  f_{\mathbf{i}}-h_{\mathbf{i}%
}(\boldsymbol{\beta})\right \}  \mid f_{i_{1}},\mathbf{X}_{i_{1}}\right]
E\left[  \mathbf{X}_{\mathbf{i}}^{\top}h_{\mathbf{i}}(\boldsymbol{\beta
})\mathbf{X}_{\mathbf{i}}\right]  ^{-1}.
\]

c) a bad (wrong) variance assumption (Constant) through\textbf{ }%
$f_{\mathbf{i}}.$ Let%

\[
V_{\mathbf{i}}=Var\left \{  f_{\mathbf{i}}\mid x_{\mathbf{i}}\right \}  =C.
\]
The optimal UGEE for estimating $\boldsymbol{\beta}$ is now%

\begin{equation}
\mathbf{U}_{n}\left(  \boldsymbol{\beta}\right)  =\sum_{\mathbf{i\in}C_{2}%
^{n}}\mathbf{D}_{\mathbf{i}}^{\top}V_{\mathbf{i}}^{-1}S_{\mathbf{i}}%
=\sum_{\mathbf{i\in}C_{2}^{n}}C^{-1}\mathbf{X}_{\mathbf{i}}^{\top
}h_{\mathbf{i}}(\boldsymbol{\beta})\left \{  f_{\mathbf{i}}-h_{\mathbf{i}%
}(\boldsymbol{\beta})\right \}  =0,
\end{equation}
yielding the asymptotic variance of%

\[
Var(\boldsymbol{\beta})=E\left[  \mathbf{X}_{\mathbf{i}}^{\top}h_{\mathbf{i}%
}(\boldsymbol{\beta})^{2}\mathbf{X}_{\mathbf{i}}\right]  ^{-1}4Var\left[
\mathbf{X}_{\mathbf{i}}^{\top}h_{\mathbf{i}}(\boldsymbol{\beta})\left \{
f_{\mathbf{i}}-h_{\mathbf{i}}(\boldsymbol{\beta})\right \}  \mid f_{i_{1}%
},\mathbf{X}_{i_{1}}\right]  E\left[  \mathbf{X}_{\mathbf{i}}^{\top
}h_{\mathbf{i}}(\boldsymbol{\beta})^{2}\mathbf{X}_{\mathbf{i}}\right]  ^{-1}.
\]
In the simulation, we used $C=\widehat{Var}\left(  f_{\mathbf{i}}\right)  .$

We set $\tau=10,\beta_{0}=3,\beta_{1}=3,a=0,b=1$ in all our similations.

\end{document}